\documentclass[aps,prd,eqsecnum,nofootinbib,twocolumn,floatfix]{revtex4}
\usepackage{graphicx}
\usepackage{times,mathptm}
\newcommand{\beq}{\begin{equation}}
\newcommand{\eeq}{\end{equation}}
\newcommand{\bea}{\begin{eqnarray}}
\newcommand{\eea}{\end{eqnarray}}

\renewcommand{\d}{\delta}
\renewcommand{\l}{\lambda}
\renewcommand{\L}{\Lambda}
\renewcommand{\b}{\beta}
\renewcommand{\a}{\alpha}

\renewcommand{\ni}{\noindent}
\newcommand{\g}{\gamma}

\newcommand{\m}{\mu}
\renewcommand{\r}{\rho}
\newcommand{\bx}{{\mathbf{x}}}
\newcommand{\by}{{\mathbf{y}}}

\newcommand{\s}{\sigma}

\newcommand{\D}{{\Delta}}

\newcommand{\R}{{\cal R}}

\newcommand{\tU}{\widetilde{U}}

\newcommand{\J}{{\cal J}}

\newcommand{\e}{\epsilon}

\newcommand{\oh}{{\textstyle{\frac{1}{2}}}}
\newcommand{\oth}{{\textstyle{\frac{1}{3}}}}

\newcommand{\dg}{\dagger}
\newcommand{\non}{\nonumber}

\newcommand{\rf}[1]{(\ref{#1})}
\newcommand{\ra}{\rightarrow}
\newcommand{\pa}{\partial}
\newcommand{\ph}{\phi}

\newcommand{\FIGURE}[2][v]{\begin{figure}[#1]#2
                                        \end{figure}}
\begin{document}
%
%
\title{Coulomb Energy, Remnant Symmetry, and the
Phases of Non-Abelian Gauge Theories}
\author{Jeff Greensite}
\affiliation{Physics and Astronomy Dept., San Francisco State
University, San Francisco, CA~94117, USA}
\email{greensit@stars.sfsu.edu}
\author{{\v S}tefan Olejn\'{\i}k}
\affiliation{Institute of Physics, Slovak Academy
of Sciences, SK--845 11 Bratislava, Slovakia}
\email{stefan.olejnik@savba.sk}
\author{Daniel Zwanziger}
\affiliation{Physics Department, New York University, New York, NY~10003, USA}
\email{daniel.zwanziger@nyu.edu}
\date{\today}
\begin{abstract}
We show that the confining property of the one-gluon
propagator, in Coulomb gauge, is linked to the unbroken
realization of a remnant gauge symmetry which exists in this
gauge. An order parameter for the remnant gauge symmetry is
introduced, and its behavior is investigated in a variety of
models via numerical simulations. We find that the color-Coulomb
potential, associated with the gluon propagator, grows linearly
with distance both in the confined and $-$ surprisingly $-$ in the
high-temperature deconfined phase of pure Yang-Mills theory. We
also find a remnant symmetry-breaking transition in SU(2)
gauge-Higgs theory which completely isolates the Higgs from the
(pseudo)confinement region of the phase diagram.  This transition
exists despite the absence, pointed out long ago by Fradkin and
Shenker, of a genuine thermodynamic phase transition separating
the two regions.
\end{abstract}
\maketitle
%
%
\section{Introduction}\label{Introduction}

    It is well known that the phase transition from a confined to
an unconfined phase in a non-abelian gauge theory is associated
with the breaking of global center symmetry. If a non-trivial
center symmetry exists and is unbroken, then Polyakov line
expectation values are zero, and in consequence the free energy of
a static color charge is infinite in an infinite volume.  But the
confined phase may be associated with other global symmetries as
well. In this article we will focus on the remnant gauge symmetry
which is found after imposing Coulomb gauge.  What is interesting
about this symmetry is that its unbroken realization implies the
existence of a confining color-Coulomb potential, and this in turn
is a necessary (but not sufficient) condition for confinement.

     The color-Coulomb potential arises from the energy of the longitudinal
color-electric field in Coulomb gauge, and corresponds
diagrammatically to instantaneous dressed one-gluon exchange
between static sources. We have previously studied this potential
numerically in pure lattice Yang--Mills theory at zero temperature,
and found that the potential rises linearly with color charge
separation \cite{DA,JS}, albeit with a string tension
$\sigma_{coul}$ which is
significantly higher (by about a factor of three \cite{JS}) than
the string tension $\sigma$ of the static quark potential. In this article
we introduce an order parameter for remnant gauge symmetry
breaking and study its behavior, as well as the behavior of the
color-Coulomb potential, in
\begin{itemize}
\item SU(2) gauge-Higgs theory, with the Higgs field in the
adjoint representation;
\item pure SU(2) gauge theory, in the confined and the
high-temperature deconfined phase;
\item SU(2) gauge-Higgs theory, with the Higgs field in the
fundamental representation;
\item compact $QED_4$.
\end{itemize}
In the first, second, and fourth theories the action has a
non-trivial global center symmetry, which may be broken
spontaneously in some range of couplings or
temperature. In the third case center symmetry is explicitly broken,
and the asymptotic string tension vanishes at all couplings. By
studying these different cases, we can explore to what extent
remnant symmetry breaking is correlated with center symmetry
breaking, and whether the confining Coulomb potential is always
associated with a confining static quark potential.

     This article is organized as follows: In section~\ref{coulomb}, below, we
introduce the remnant symmetry order parameter, and relate it to
the confining properties of the color-Coulomb potential. We also
examine scaling of $\s_{coul}$ with $\b$. Section~\ref{gauge-adjoint-higgs-1}
concerns the gauge-adjoint Higgs
model, where we find perfect correspondence between remnant
symmetry and center symmetry breaking.  But this correspondence is
lost already in pure Yang--Mills theory at high temperature,
studied in section~\ref{highT}, where we find that Coulomb
confinement and unbroken remnant symmetry persist in the
deconfined phase.  A possible explanation of this phenomenon is
discussed. In section~\ref{confinement-scenario} we review the
Gribov confinement scenario in Coulomb gauge and show that it
accords with the vortex dominance scenario, by gauge transforming
from the maximal center gauge to the minimal Coulomb gauge. In
section~\ref{gauge-fundamental-higgs} we present our results for
the gauge-fundamental Higgs model, where the gauge-Higgs
interaction breaks global center symmetry explicitly. In this case
we find very clear numerical evidence of a remnant symmetry-breaking
transition which is unaccompanied by a true thermodynamic phase
transition, and also argue for the existence of such a transition
from a lattice strong-coupling
analysis. This result is in complete accord with the earlier
work of Langfeld \cite{Kurt1,Kurt2}, which found remnant symmetry
breaking in Landau gauge, and it implies that a sharp distinction can
be made between the Higgs and the (pseudo)confining regions of the
gauge-Higgs coupling plane. This distinction exists despite the
fact, pointed out by Fradkin and Shenker \cite{FS}, that these
regions are continuously connected in the usual sense of
thermodynamics.  The remnant symmetry breaking in the adjoint
Higgs theory is re-examined in section~\ref{gauge-adjoint-higgs-2},
where we draw some conclusions about the measure of abelian configurations
in the fundamental modular region.
Our methods are applied to compact $QED_4$ in
section~\ref{qed}.

   As in the previous work of ref.\ \cite{JS}, we also investigate
numerically the relevance of center vortices to the existence of a
confining Coulomb potential, particularly in the high-temperature
deconfined phase of pure Yang--Mills theory, and in the
(pseudo)confined phase of gauge-fundamental Higgs theory.  In
section~\ref{confinement-scenario} it is shown that thin center
vortices lie on the Gribov Horizon, which may be relevant to their
dramatic effect on the Coulomb potential.  Vortex removal, by the
de Forcrand--D'Elia procedure \cite{dFE}, is found in every case
to convert a confining Coulomb potential to a non-confining,
asymptotically flat potential.

    In section~\ref{translation}, we provide a 2-way translation between the temporal gauge
($A_0 = 0$) and the minimal Coulomb gauge.  This allows our measurements,
which are made by gauge-fixing to the minimal Coulomb gauge, to be
equivalently described in temporal gauge.  We show that the state in
which our measurements are made is, in temporal gauge,
a quark-pair state of the
type introduced by Lavelle and McMullan \cite{lavelle}.  It
has correct gauge-transformation properties, although it does not
make use of Wilson lines running between the sources.
Section~\ref{conclusions} contains some concluding remarks.

%
%
\section{Coulomb Energy and Remnant Symmetry}\label{coulomb}

     On the lattice, minimal Coulomb  gauge consists of fixing to the
configuration on the gauge orbit maximizing the quantity
\beq
        R = \sum_{\bx,t} \sum_{k=1}^3 \mbox{Tr}[U_k(\bx,t)].
\eeq
\ni Maximizing $R$ does not fix the gauge completely, since there
is still the freedom to perform time-dependent gauge
transformations
\bea
      U_k(\bx,t) &\ra& g(t) U_k(\bx,t) g^\dg(t) \qquad (k=1,2,3),
\non \\
      U_0(\bx,t) &\ra& g(t) U_0(\bx,t) g^\dg(t+1).
\eea
To understand the role of this remnant gauge symmetry,
consider a state $\Psi_q^a$ obtained by operating on the lattice
Yang--Mills vacuum state $\Psi_0$ (i.e.\ the ground state of the
Coulomb gauge transfer matrix) with a heavy quark operator $q^a$
\beq
     \Psi_q^a[\bx;A] = q^a(\bx) \Psi_0[A].
\label{psia}
\eeq
Evolve this state for Euclidean time $T<n_t$, where $n_t$ is the
lattice extension in the time direction. Dividing out the vacuum factor
$\exp[-E_0 T]$ and taking the
inner-product with $\Psi_q^b[\bx;A]$, we have
\bea
        G^{ba}(T) &=& \langle \Psi_q^b | e^{-(H-E_0)T} | \Psi_q^a \rangle
\non \\
    &=& \sum_n \langle \Psi_q^b |\Psi_n \rangle \langle \Psi_n |
\Psi_q^a \rangle
          e^{-(E_n - E_0) T}
\non \\
    &=&  \Gamma e^{-m_q T}  \langle L^{ba}(\bx,1,T) \rangle
\eea
where index $n$ refers to summation over charged energy
eigenstates, $m_q \ra \infty$ is the heavy quark mass, $\Gamma$ is
a numerical factor, and $L(\bx,t_1,t_2)$ is a Wilson line
\beq
       L(\bx,t_1,t_2) = U_0(\bx,t_1) U_0(\bx,t_1+1) \dots U_0(\bx,t_2).
\eeq

    In a confining theory, the excitation energy $\D E_n = E_n-E_0$ of any
state $\Psi_n$  containing a single static quark must be infinite.
It follows that $G^{ba}(T)=0$, and therefore that the expectation
value of any timelike Wilson line $L(\bx,1,T)$ must vanish.  Now,
under the remnant gauge symmetry, $L$ transforms as
\beq
       L(\bx,t_1,t_2) \rightarrow g(t_1)L(\bx,t_1,t_2)g^\dg(t_2+1).
\eeq \ni  If $T=t_2-t_1+1$ is less than the lattice extension
$n_t$ in the
time direction, then not only $L$ but also $\mbox{Tr}[L]$ is
non-invariant under the remnant gauge symmetry.\protect\footnote{If
$T=n_t$, then $L$ is a Polyakov line, whose trace is invariant
under gauge transformations, but non-invariant under global center
transformations.} If the remnant symmetry is unbroken, then
$\langle L \rangle$ must vanish.  The confining phase is therefore
a phase of unbroken remnant gauge symmetry; i.e.\ unbroken remnant
symmetry is a necessary condition for confinement.

    There are several issues which require some further
comment.  First of all, why is unbroken remnant symmetry not also
a \emph{sufficient} condition for confinement?  The answer is that
in an unconfined phase, where there exist finite energy states
containing a single static charge in an infinite volume, there is
still the possibility that these finite energy states have
vanishing overlap with $\Psi_q^a$ as defined in eq.\ \rf{psia}.
Thus $\langle L \rangle = 0$ and unbroken remnant symmetry could
be found, in principle, also in the absence of confinement.
Secondly, it is obviously
impossible to insert a single charge in a finite volume with,
e.g., periodic boundary conditions; electric field lines starting
from the static charge must end on some other charge, regardless
of whether or not the theory is in a confining phase. How, then,
is this fact reflected in the remnant symmetry-breaking criterion?
The same question can be raised in connection with
Polyakov lines, and the answer is the same: Strictly speaking,
spontaneous symmetry breaking cannot occur in a finite volume, so
$\langle L \rangle = 0$ always, consistent with the absence of an
isolated static charge. Nevertheless (again like Polyakov lines),
it is possible to construct an order parameter which detects the
infinite-volume transition via finite-volume calculations which
are subsequently extrapolated to infinite volume.\protect\footnote{In the
case of spontaneous center symmetry breaking, the accepted order
parameter on a finite lattice is the absolute value of the spatial
average of Polyakov lines.} We will construct such an operator
below. Finally, there is the question of Elitzur's theorem.
Although the remnant symmetry is global on a time slice, it is
local in the time direction, and according to the theorem local
symmetries cannot break spontaneously.  So how could we ever have
$\langle L \rangle\ne 0$, even in an infinite volume?  The answer
is that in fact the average value of $L(\bx,t,T+t)$ does indeed
vanish on an infinite lattice, in accordance with the Elitzur
theorem, providing the averaging is done over all spatial $\bx$
and all times $t$. On a time slice, however, the symmetry is only
global, and it is possible in any given configuration that the
average value of $L(\bx,t,T+t)$ is finite on an infinite lattice,
when averaged over all $\bx$ at \emph{fixed} time~$t$. This is
what we will mean by the phrase ``spontaneous breaking of the
remnant symmetry," and it involves no actual violation of the
Elitzur theorem.

    With these points in mind, we propose to construct an order parameter
for remnant symmetry breaking from the timelike link variable
averaged over spatial volume at fixed time.  Let
\beq
       \tU(t) = {1\over V_3} \sum_{\bx} U_0(\bx,t)
\eeq \ni where $V_3=n_x n_y n_z$ is the is the 3-volume of a
lattice time-slice. If remnant symmetry is unbroken, then $\tU(t)
= 0 + O(1/V_3^{1/2})$ in any thermalized lattice configuration.
The order parameter $Q$ is defined to be, for SU(2),
\beq
     Q = {1\over n_t} \sum_{t=1}^{n_t} \left\langle \sqrt{\oh \mbox{Tr}[\tU(t)
     \tU^\dg(t)]} \right\rangle.
\label{Q}
\eeq
\ni Then $Q$ is positive definite on a finite
lattice, and on general grounds \beq
       Q = c + {b \over \sqrt{V_3}}  ~~~\mbox{where}~~
         \left\{ \begin{array}{cl}
                   c  = 0 & \mbox{~ in the symmetric phase,} \cr
                   c  > 0 & \mbox{~ in the broken phase.} \end{array}
              \right.
\eeq
\ni If $Q$ extrapolates to a non-zero value as $V_3\ra \infty$, then
the remnant symmetry in Coulomb gauge is spontaneously broken.

     Next, we make the connection between unbroken remnant symmetry and the
existence of a confining Coulomb potential. We first recall that
the Hamiltonian operator in Coulomb gauge has the form $H = H_{glue} +
H_{coul}$ where, in the continuum
\bea
       H_{glue} = \oh \int d^3x\;( \J^{-\oh}\vec{E}^{{\rm tr},a} {\cal J}
       \cdot \vec{E}^{{\rm tr},a} \J^{-\oh} + \vec{B}^a \cdot \vec{B}^a)
\non \\
       H_{coul} = \oh \int d^3x d^3y\;\J^{-\oh}\r^a(x) \J
               K^{ab}(x,y;A) \r^b(y) \J^{-\oh}
\non
\eea
\vspace*{-5mm}
\bea
       K^{ab}(x,y;A) &=& \left[ {1\over \nabla \cdot D(A)}
       (-\nabla^2) {1\over \nabla \cdot D(A)} \right]^{ab}_{xy}
\non \\
       \r^a &=& \r_q^a - g f^{abc} A^b_k E^{{\rm tr},c}_k
\non \\
        \J &=& \det[-\nabla \cdot D(A)]
\eea
and the factors of $\J$ arise from operator ordering considerations
\cite{Christ}.  It is understood that $A$ is identically transverse
in Coulomb gauge, $A = A^{\rm tr}$.  Note that $\J$ commutes
with all quantities except $E^{\rm tr}$, and in particular with
$\r_q(x)$ and $K^{ab}(x,y;A)$.
The expectation-value of $K(x,y;A)$ is the instantaneous piece
of the $\langle A_0 A_0 \rangle$ gluon propagator, i.e.
\bea
         \langle A^a_0(x) A^b_0(y) \rangle &=&
D(\bx-\by)\d^{ab} \d(x_0-y_0) \non \\
            &+& \mbox{non-instantaneous}
\non \\
      D(\bx-\by)\d^{ab} &=& \left\langle \left[{1\over \nabla
\cdot D[A]}
    (-\nabla^2) {1\over \nabla \cdot D[A]} \right]_{x,y}^{a,b} \right\rangle,\non \\
& &
\label{D}
\eea
\ni as shown in \cite{rengrcoul}.
We see from these expressions that the Coulomb interaction
energy between two charged static sources is given by instantaneous
(dressed) one-gluon exchange.

    Now consider a physical state in Coulomb gauge containing massive
quark-antiquark sources
\beq
    |\Psi_{qq}\rangle = \overline{q}(0) q(R) |\Psi_0 \rangle
\eeq
which is invariant under the remnant symmetry.  The
excitation energy is
\bea
        {\cal E} &=& \langle \Psi_{qq}|H|\Psi_{qq}\rangle
                 - \langle \Psi_0|H|\Psi_0\rangle
\non \\
                 &=& V_{coul}(R) + E_{se},
\eea
where $E_{se}$ is an $R$-independent constant, on the order of the
inverse lattice spacing, to be specified below.
The $R$-dependence of ${\cal E}$
can only come from the expectation value of the non-local quark-quark
part of the Hamiltonian
\beq
   H_{qq} = \oh \int d^3x d^3y \ \r_q^a(x) \ K^{ab}(x, y; A) \ \r_q^b(y).
\eeq
\ni Thus the $R$-dependent piece $V_{coul}(R)$ can be
identified as the Coulomb potential due to these static sources.
Moreover, the same kernel $K(x, y; A)$ appears in $H_{qq}$ and
in the instantaneous part $D(\bx)$
of the (dressed) one-gluon propagator $\langle A_0 A_0 \rangle $.
This yields the formula,
\bea
        V_{coul}(|\bx|) + E_{se} &=& C_r \Bigl( D(0) - D(\bx) \Bigr),
\label{v-prop}
\eea
where $C_r$ ($=3/4$ for the SU(2) gauge group) is the Casimir
factor in the fundamental representation.

     The correlator of two Wilson lines can be expressed in terms
of the Hamiltonian operator and the state $\Psi_{qq}$ as follows:
\bea
     G(R,T) &=& \langle \oh\mbox{Tr}[L^\dg(\bx,0,T) L(\by,0,T)] \rangle
\non \\ \non \\
            &=& \langle \Psi_{qq} | e^{-(H-E_0)T} | \Psi_{qq}
            \rangle,
\eea
\ni where $R=|\bx-\by|$, and $L$ is now the timelike Wilson line in
the continuum theory.   We have
\beq
   G(R,T) = \sum_n \left| \langle \Psi_n|\Psi_{qq}\rangle \right|^2
e^{-\D E_n T}
\eeq
\ni and we define the logarithmic derivative
\beq
V(R,T) = -{d\over dT} \log[G(R,T)].
\label{vrt}
\eeq
It is easy to see that the
Coulomb energy is obtained at $T\ra 0$, i.e.
\bea
      {\cal E} &=& V_{coul}(R) + E_{se}
\non \\
               &=& V(R,0).
\label{V1}
\eea
\ni The minimum energy of a state containing two
static quark antiquark charges, which in a confining theory would
be the energy of the flux tube ground state, is obtained in the
opposite $T\ra \infty$ limit
\bea
      {\cal E}_{min} &=& V(R) + E'_{se}
\non \\
                     &=& \lim_{T\ra \infty} V(R,T)
\label{V2}
\eea
\ni and in this limit $V(R)$ is usual static quark potential.

     The idea of using the correlator of timelike Wilson lines
to compute the static quark potential, in Coulomb gauge on the
lattice, was put forward some years ago by Marinari et al.\
\cite{Marinari}. These authors also noted that the remnant
symmetry in Coulomb gauge is unbroken in the confining phase.

    We now recall an inequality first pointed out by one of us
(D.Z.) in ref.\ \cite{Dan}.  With a lattice regularization,
$E_{se}$ and $E'_{se}$, are finite constants.  In a
confining theory, both of these constants are negligible compared
to $V(R)$, at sufficiently large $R$.  But since ${\cal E}_{min}
\le {\cal E}$, it follows that
\beq
           V(R) \le V_{coul}(R)
\eeq
\ni asymptotically.  The intriguing implication is that if
confinement exists at all, then it exists already at the level of
dressed one-gluon exchange in Coulomb gauge.  But we also see that
because the Coulomb potential is only an upper bound on the static
potential, a confining Coulomb potential is a necessary but not a
sufficient condition for the existence of a confining static quark
potential.

    On the lattice, the continuum logarithmic derivative in eq.\
\rf{vrt} is replaced by
\beq
   V(R,T) = {1\over a} \log\left[G(R,T) \over G(R,T+a) \right]
\eeq
\ni where $a$ is the lattice spacing.  In particular, in lattice
units $a=1$,
\bea
   V(R,0) &=& -\log[G(R,1)]
\non \\
          &=& -\log\left[\left\langle \oh\mbox{Tr}[U_0(\bx,1)
          U_0^\dg(\by,1)]\right\rangle \right],
\label{vr0}
\eea
\ni and at large $\b$, where the lattice logarithmic derivative
approximates the continuum, $V(R,0)$ provides an estimate of the
Coulomb potential $V_{coul}(R)$ (up to an additive constant $E_{se}$).

    In eq.\ \rf{vr0} the relation between the confining property
of the Coulomb potential, and the unbroken realization of remnant
symmetry, is manifest.  For if $Q\ra 0$ at infinite volume, then
also
\bea
\lim_{R\ra \infty}\langle
\mbox{Tr}[U_0(\bx,t)U_0^\dg(\by,t)]\rangle &=&  \lim_{V_3\ra
\infty}\langle \mbox{Tr}[\tU(t)\tU^\dg(t)]\rangle
\non \\
    &=& 0
\label{limit}
\eea
\ni in which case the potential $V(R,0)$ rises
to infinity as $R\ra \infty$. Conversely, if $Q>0$, then the limit
in eq.\ \rf{limit} is finite, and  $V(R,0)$ is asymptotically
flat. Since $V_{coul}(R)\approx V(R,0)$ is an upper bound on the
static quark potential, we see again that unbroken remnant
symmetry is a necessary but not sufficient condition for
confinement.

     $V(R,T)$ has been computed numerically
for pure SU(2) lattice gauge theory at
a range of lattice couplings $\b$ in ref.\ \cite{JS}.  We recall
the essential results:
\begin{enumerate}
\item $V(R,T)$ increases linearly with $R$ at large $R$ and all
$T$, at any coupling.
\item The associated string tension $\s(T)$
converges (from above) to the usual asymptotic string tension $\s$, at
any given $\b$, as $T$ increases.
\item At weaker couplings, the
Coulomb string tension $\s_{coul} \equiv \s(0)$ is substantially
greater (by about a factor of three) than the asymptotic string
tension.
\item Removing center vortices from lattice
configurations (by the de Forcrand--D'Elia procedure \cite{dFE})
sends $\s(T) \ra 0$ at all $T$, including the Coulomb string
tension $\s_{coul}\ra 0$ at $T=0$.
\end{enumerate}

In the following sections we extend the investigation to models
including scalar matter fields, and to SU(2) gauge theory across
the high temperature deconfinement transition.\protect\footnote{A note on
gauge fixing to Coulomb gauge:  In this investigation we generate eight
random gauge copies of each lattice configuration, and carry out
gauge-fixing on each configuration by over-relaxation for 250 iterations.
The best copy of eight copies is then chosen, and the over-relaxation
procedure is continued  until the average value of the
gauge-fixed links has changed, in the last 10 iterations, by less than
$2\times 10^{-7}$.} First, however, we
would like to remark on the scaling properties of $\s_{coul}
\approx \s(0)$ in pure SU(2) gauge theory at zero temperature. The
ratio of the Coulomb ($T=0$) to the asymptotic ($T\ra \infty$)
string tensions, reported in ref.\ \cite{JS}, varies somewhat with
$\b$ in the range ($\b\in [2.2,2.5]$) of couplings investigated.
The ratio $\s(0)/\s$ tends to rise in this interval, as shown in
Fig.\ \ref{sigmavsbeta}.  However, it is known that $\s$ does not
quite conform to the two-loop scaling formula associated with asymptotic
freedom, in the range of $\b$ we have studied, and it is
always possible that scaling sets in at different values of $\b$
for different physical quantities. What we find is that when our
values for $\s(0)$ are divided by the asymptotic freedom
expression
\beq
       F(\b) = \left( {6\pi^2 \over 11} \b \right)^{102/121}
                  \exp\left(-{6\pi^2 \over 11} \b \right)
\eeq
relevant to the SU(2) string tension, the ratio $\s(0)/F(\b)$
is virtually constant as seen in Fig.\ \ref{sigmavsbeta3}. This fact
suggests that scaling according to asymptotic freedom may set in
earlier for the Coulomb string tension $\s_{coul}\approx \s(0)$
than for the asymptotic string tension $\s$.

\FIGURE[tb]{
\centerline{{\includegraphics[width=8truecm]{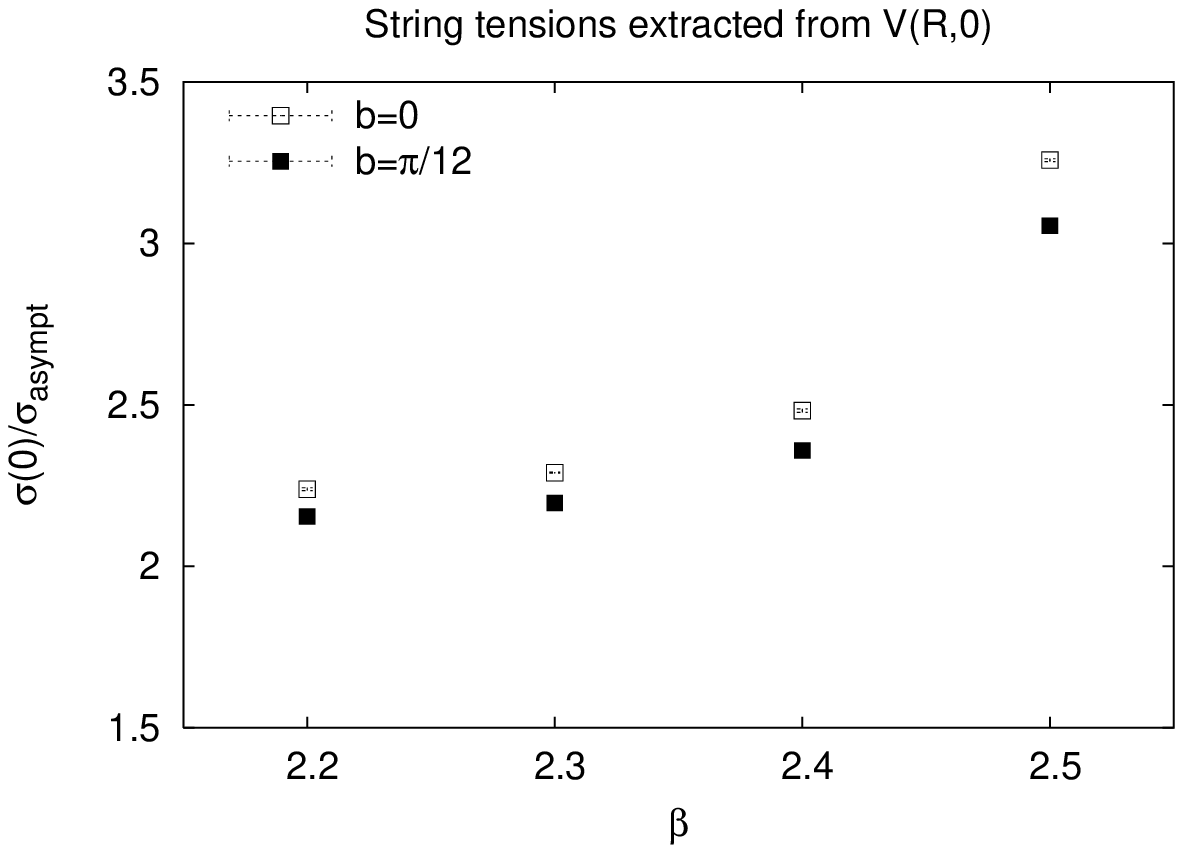}}}
\caption{The ratio $\s(0)/\s$ at various $\beta$, from fits which
either include ($b=\pi/12$) or do not include ($b=0$) the
L\"uscher term.  This ratio should equal the ratio of
$\s_{coul}/\s$ in the continuum limit. (Data points come from numerical
simulations on lattice sizes $16^4$, $20^4$, $20^4$, and $32^4$ at
$\beta=2.2, 2.3, 2.4,$ and $2.5$ respectively.)}
\label{sigmavsbeta} }

\FIGURE[tb]{
\centerline{{\includegraphics[width=8truecm]{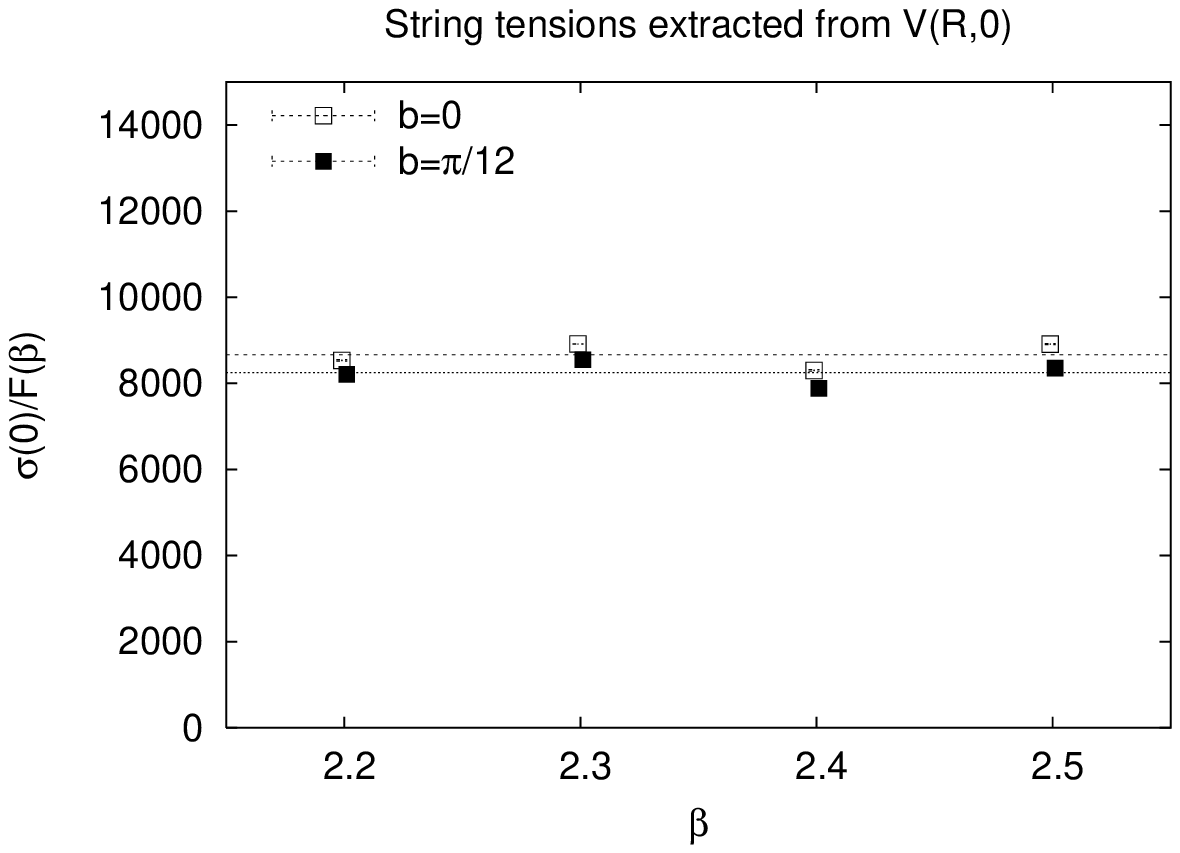}}}
\caption{The ratio $\s(0)/F(\b)$ vs.\ $\b$, again from fits
with ($b=\pi/12$) or without ($b=0$) the
L\"uscher term.  $F(\b)$ is the asymptotic freedom expression
given in the text; a constant ratio implies scaling according to
the two-loop beta function.}
\label{sigmavsbeta3} }

\subsection{Divergent constant in $D(R)$}

  We would also like to remark, at this stage, on a subtlety in
identifying the color Coulomb potential with dressed one-gluon
exchange.\footnote{We thank M. Polikarpov for a helpful discussion
on this point.} The energy expectation value of the static
$q\overline{q}$ state,
${\cal E}(R) = V_{coul}(R) + E_{se}$, is finite for finite
quark separation $R$ and finite lattice spacing $a$; in fact we have calculated
this quantity ($=V(R,0)$) numerically.  So it might seem natural, from eq.\
\rf{v-prop}, to identify the $R$-dependent Coulomb interaction
energy as $V_{coul}(R)= - C_r D(R)$. That cannot be quite right,
however. The reason is that $\oh C_r D(0)$ is the energy of an
\emph{isolated} quark state in an infinite volume; it is the
energy we would extract from the logarithmic time derivative of
$G(T)=\langle \oh \mbox{Tr}L(0,0,T)\rangle$. In the case of
unbroken remnant symmetry we have $G(T)=0$, and therefore $\oh C_r
D(0) = \infty$ in an infinite volume, even though the lattice
spacing $a$ is non-zero.  This infinity, which has a
non-perturbative, infrared origin, should not be confused with the
usual ultraviolet contribution to the quark self-energy, which is
only infinite in the continuum limit. Since ${\cal E}(R)$ is finite,
the infrared divergence in $C_r D(0)$ must
be cancelled, in eq.\ \rf{v-prop}, by a corresponding
divergent constant contained in $C_r D(R)$.  In other words, only
the difference $D(0) - D(R)$ is finite, and $V_{coul}(R)$, if finite,
differs from $- C_r D(R)$ by an infinite constant.   In order that
$V_{coul}(R)$ and $E_{se}$ be separately finite and well-defined,
we may relate them to the gluon propagator with an (arbitrary)
subtraction at $R=R_0$ which removes the infrared divergence, i.e.
\bea
        V_{coul}(R) &=&  - C_r\Bigl( D(R) - D(R_0) \Bigl)
\non \\
                    &=& {\cal E}(R) -  {\cal E}(R_0)
\non \\
         E_{se} &=& C_r\Bigl( D(0) - D(R_0) \Bigl)
\non \\
                &=& {\cal E}(R_0)
\eea
Defined in this way, $V_{coul}(R)$ crosses zero at the subtraction point,
and $E_{se}$ contains the ultraviolet, but not the infrared,
contributions to the quark-antiquark self-energies.

   As a check of the cancellation (or non-cancellation) of infrared
divergences, consider a colored state consisting, e.g., of two
static quarks, rather than a quark and antiquark. The energy of
such a state could be extracted from an $LL$ correlator, which is
zero if remnant symmetry is unbroken. The energy is therefore
infinite, and according to our previous analysis would be
proportional to $D(0)+D(R)$.  In this case, the divergent constant
in $D(R)$ adds to, rather than subtracts from, the divergent
constant in $D(0)$, and the resulting sum is divergent, as it
should be. The argument can be readily generalized to baryonic
states in SU($N$) gauge theories composed of static charges.  The
energy of a color singlet state, with charges at points $\bx_1,
\bx_2, \dots, \bx_N$ is obtained from the logarithmic time
derivative of the correlator \beq
   G(\{\bx_i\},T) = \epsilon_{i_1\dots i_N} \epsilon_{j_1\dots j_N}
     \langle L^{i_1 j_1}(\bx_1,0,T)\dots L^{i_N j_N}(\bx_N,0,T) \rangle.
\eeq
The order $T$ contributions to $G(\{\bx_i\},T)$ are terms
proportional to $D(0)$ and $D(\bx_m-\bx_n),~m\ne n$, with
differing signs.  On the other hand, for a color singlet state,
the operator $\e \e\;L\dots L$ is a $T$-independent constant in
the $x_1=x_2=\dots=x_N$ coincidence limit.  From this it is clear
that the propagators completely cancel in the coincidence limit,
and any constant terms in the propagators cancel in general.  This
means that the energy of a color singlet baryonic state is finite.
Conversely, the divergent constants do not cancel in color
non-singlet states, so their energies are infinite.

%
%
\section{SU(2) Gauge-Adjoint Higgs Theory}\label{gauge-adjoint-higgs-1}

    The lattice action for SU(2) gauge theory with a Higgs field
in the adjoint representation of the gauge group is
\bea
    S &=& \b \sum_{plaq} \oh \mbox{Tr}[UUU^\dg U^\dg] \\
  &+&\frac{\gamma}{4} \sum_{x,\m} \phi^a(x)\phi^b(x+\widehat{\m})
          \mbox{Tr}[\s^a U_\m(x) \s^b U_\m^\dg(x)] \non
\label{adjhiggs}
\eea
with the radially ``frozen", three-component Higgs field $\phi$
subject to the restriction
\beq
           \sum_{a=1}^3 \phi^a(x) \phi^a(x) = 1.
\eeq
This theory was first studied numerically by Brower et al.\
\cite{Brower}.  In addition to SU(2) gauge symmetry, the
action is also invariant under global $Z_2$ center symmetry
\beq
         U_0(\bx,t_0) \ra  - U_0(\bx,t_0)
\label{center}
\eeq
for some choice of $t=t_0$, with all other fields unchanged. The
existence of this apparently innocent global symmetry in the
action has profound consequences; in its absence the static
quark potential is asymptotically flat, and there can be no truly
confined phase.  The significance of center
symmetry to the confinement property in general is reviewed in
ref.\ \cite{review}.

\FIGURE[tb]{
\centerline{~\hspace*{-10mm}{\includegraphics[width=8truecm]{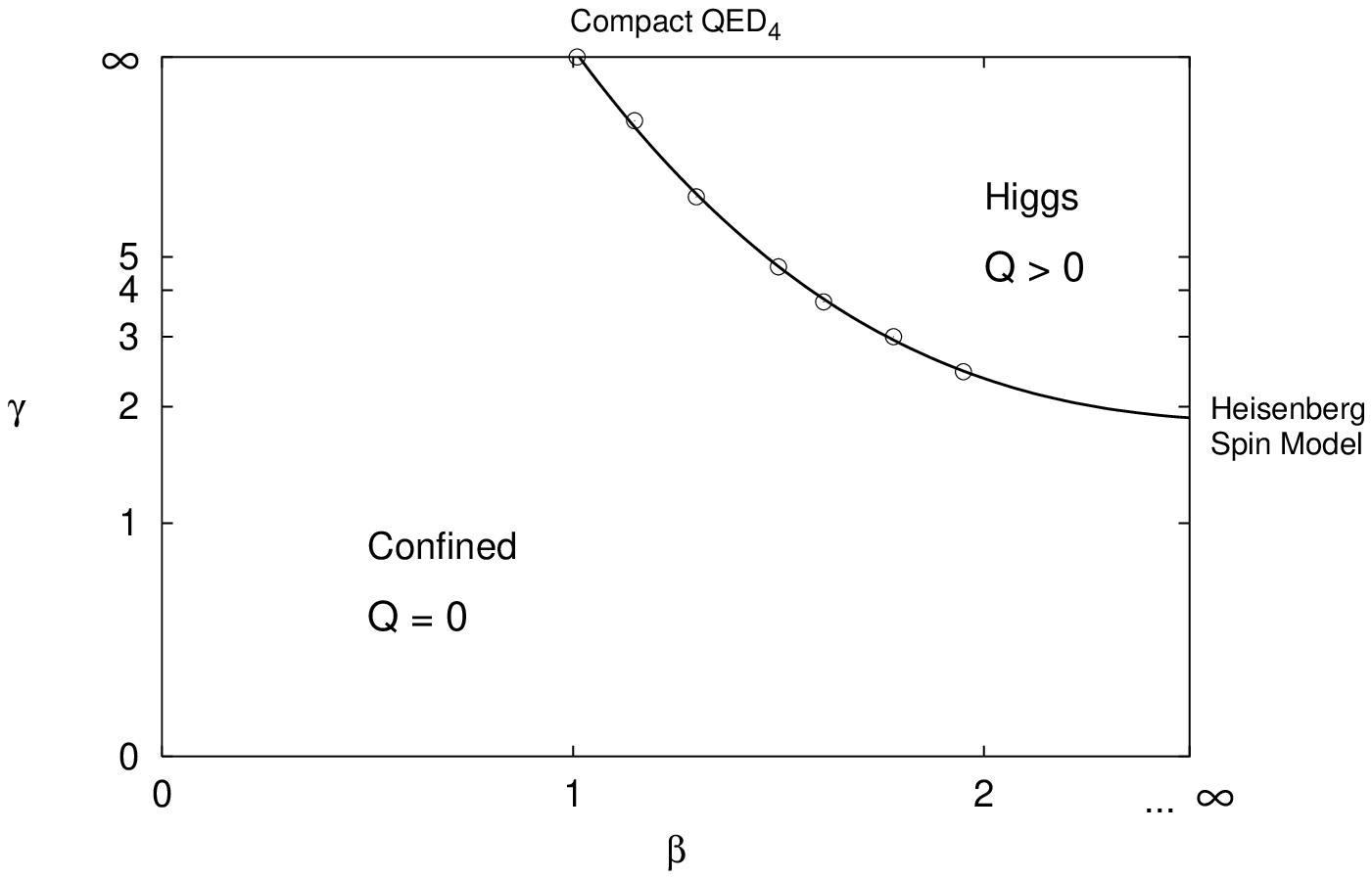}}}
\caption{Phase diagram of the SU(2) adjoint Higgs model.  The
plaquette energy $E_p$ and the remnant symmetry order parameter
$Q$ locate the same transition line between the confined and
Higgs phases.}
\label{phase_a}
}

    Since the action \rf{adjhiggs} is symmetric under
global center transformations, a transition from the Higgs phase
to a distinct confinement phase is possible.  The Higgs phase is
the phase of spontaneously broken center symmetry, while
confinement corresponds to the symmetric phase.  This division of
the $\b-\gamma$ phase diagram into two separate phases was
verified numerically long ago, in the Monte Carlo investigation of
ref.\ \cite{Brower}, which mapped out the approximate location of
the transition line.

\FIGURE[tb]{
\centerline{{\includegraphics[width=8truecm]{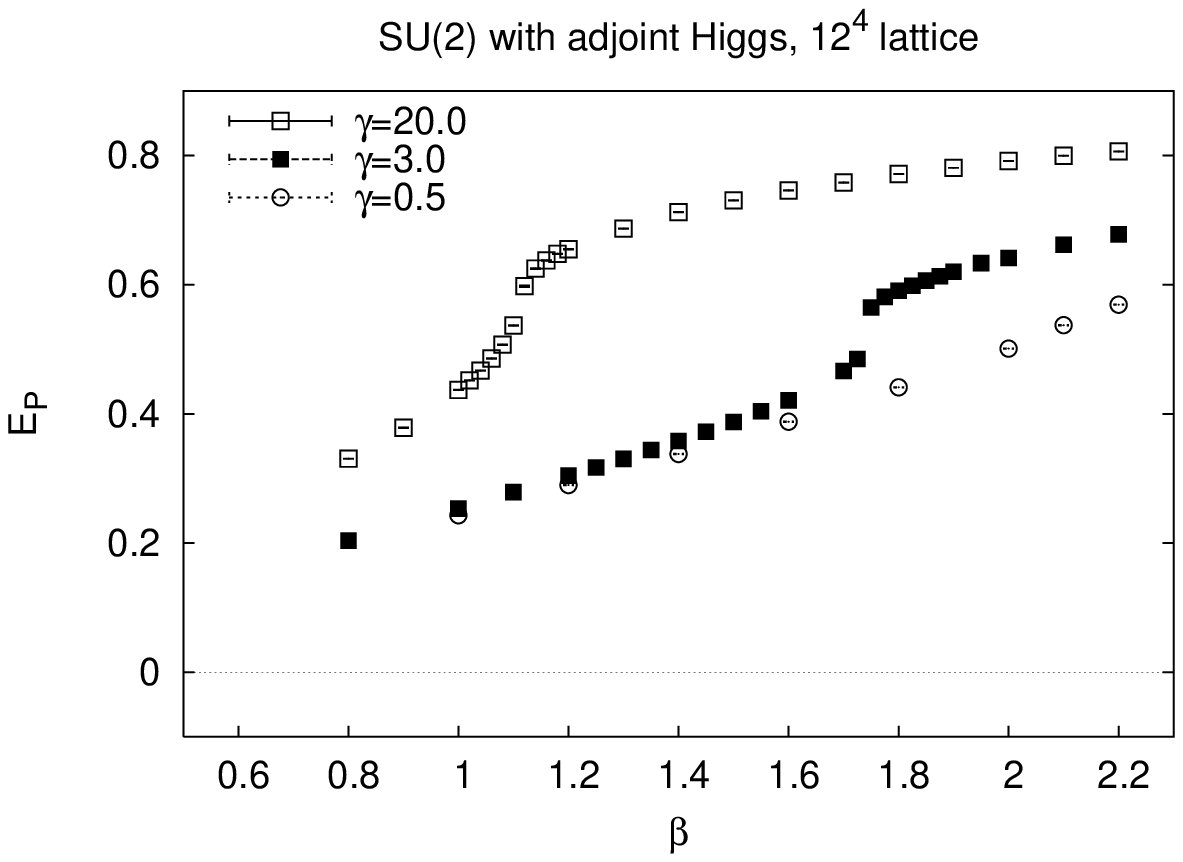}}}
\caption{Plaquette energy $E_p$ vs.\ $\b$ at three values of
$\gamma$ in the gauge-adjoint Higgs model.}
\label{ep12}
}

\FIGURE[tb]{
\centerline{{\includegraphics[width=8truecm]{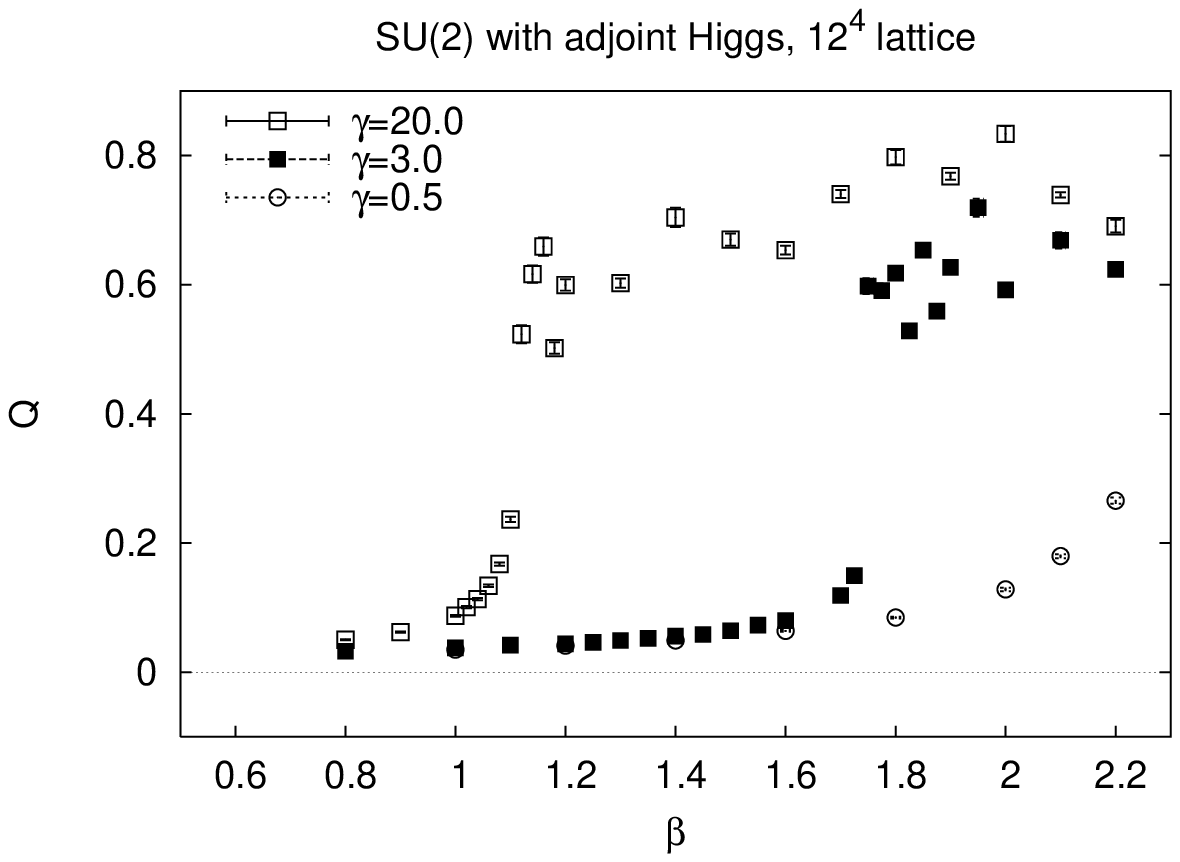}}}
\caption{Remnant symmetry order parameter $Q$ vs.\ $\b$ at
three values of $\gamma$ in the gauge-adjoint Higgs model.}
\label{q12}
}

    The Higgs phase of the adjoint Higgs model is often characterized
as a spontaneous breaking of the local gauge symmetry, from
$SU(2)$ down to $U(1)$.  In view of the Elitzur theorem, which
states that a local symmetry cannot break spontaneously, this
characterization is a little misleading.  However, as we have discussed
above, there exists in Coulomb
gauge a remnant gauge symmetry which is global
on a time slice, and which can break spontaneously on a time slice
in the sense described in the previous section.  We have therefore
studied the phase diagram of the adjoint Higgs theory via two
observables:  (i) the plaquette energy

\beq
        E_p = \Bigl\langle \oh \mbox{Tr}[UUU^\dg U^\dg] \Bigr\rangle
\eeq

\ni and (ii) the remnant symmetry breaking order parameter $Q$ defined
in eq.\ \rf{Q}.  What we find is that the transition
lines (Fig.\ \ref{phase_a}) detected by each of these two parameters
coincide; the common line location
agrees with the earlier results of Brower et al.\ based on the
plaquette energy alone.  In Fig.\ \ref{ep12} we plot $E_p$ vs.\ $\b$ at
three values of $\g$.  The existence of a phase transition for the
two larger values of $\g$ is clearly visible; there is no
transition apparent at the smallest $\g$ value.  Fig.\ \ref{q12} is the
corresponding plot of $Q$ vs.\ $\b$ at the same three values of
$\g$.  At $\b,\g$ values where $E_p$ shows a transition, the
transition in $Q$ is even more evident.  Conversely, where no
transition is seen in $E_p$, at $\g=0.5$, neither is there a
transition in $Q$. Finally, in Fig.~\ref{qvsL},
$Q$ is plotted against $V_3^{-1/2}$, and
we show the extrapolation
of $Q$ to a small value (consistent with zero) at infinite volume, for
couplings in the confined phase.

\FIGURE[tb]{
\centerline{{\includegraphics[width=8truecm]{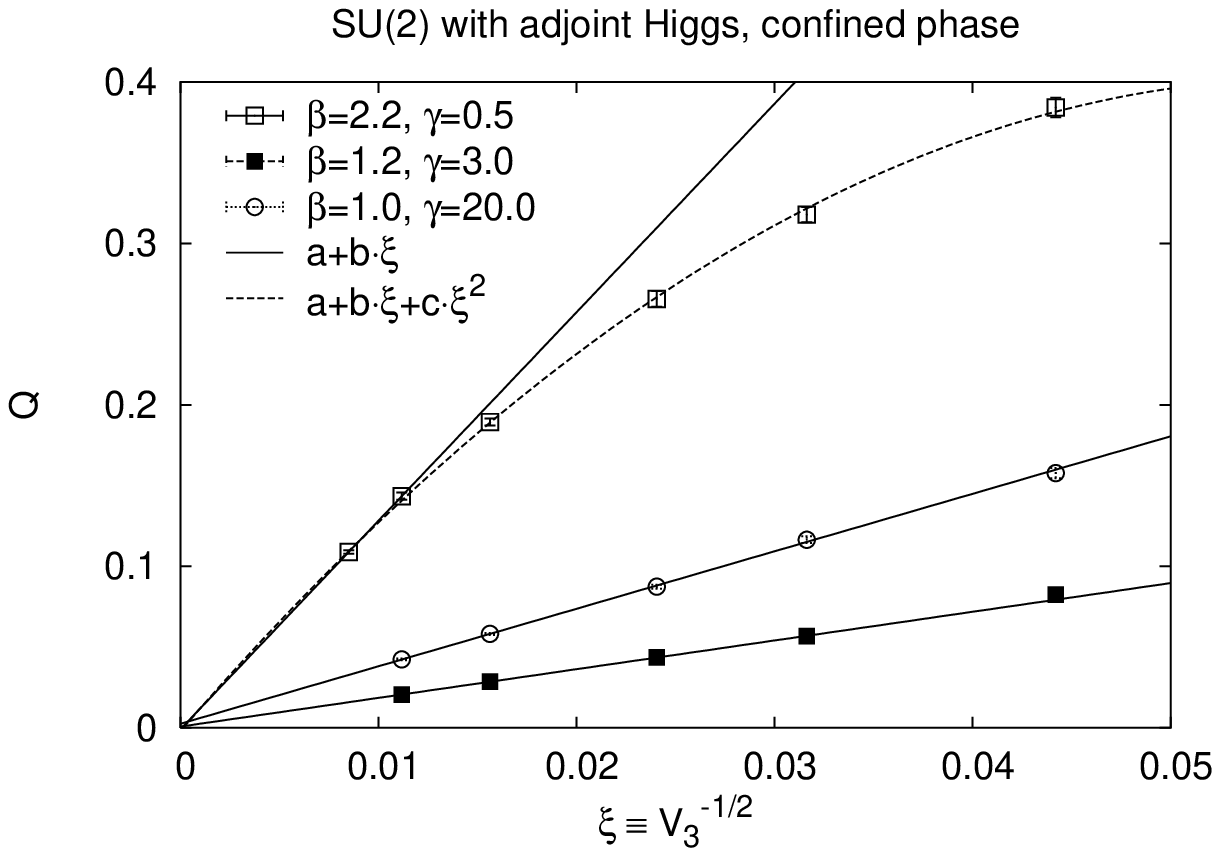}}}
\caption{Plot of $Q$ vs.\ root inverse 3-volume, and
extrapolation of $Q$ to infinite volume, at several
couplings in the confined phase of the gauge-adjoint Higgs model.}
\label{qvsL}
}

%
%
\section{High-Temperature Deconfinement}\label{highT}

    We have seen that in the SU(2)-adjoint Higgs model, things go
much as one might have expected a priori:  remnant symmetry
breaking coincides with $Z_2$ center symmetry breaking, and in
consequence the presence of a confining Coulomb potential is
correlated with the presence of a confining static quark
potential.  One might then guess that remnant and center symmetry
breaking always go together.  This appears not to be true, as we
have discovered in our investigation of pure SU(2) lattice gauge
theory at high temperature.

    Monte Carlo simulations of the pure SU(2) gauge theory were
carried out on $L^3 \times 2$ lattices at $\b=2.3$, which is inside the
deconfined phase.  Figure \ref{q1} is a plot of $Q$ vs.\
$1/\sqrt{V_3}$, where it seems that $Q$ tends to zero at large
volume (although we cannot entirely rule out a small non-zero
intercept at $V_3 \ra \infty$).  This impression is strengthened
by our plot of $V(R,0)$ in Fig.\ \ref{v1}, where it is clear that the
Coulomb potential goes asymptotically to a straight line, as the
lattice volume increases.  There is no indication of the screening
of the static quark potential (as measured by Polyakov line
correlators), which occurs at much smaller distances.

\FIGURE[tb]{
\centerline{{\includegraphics[width=8truecm]{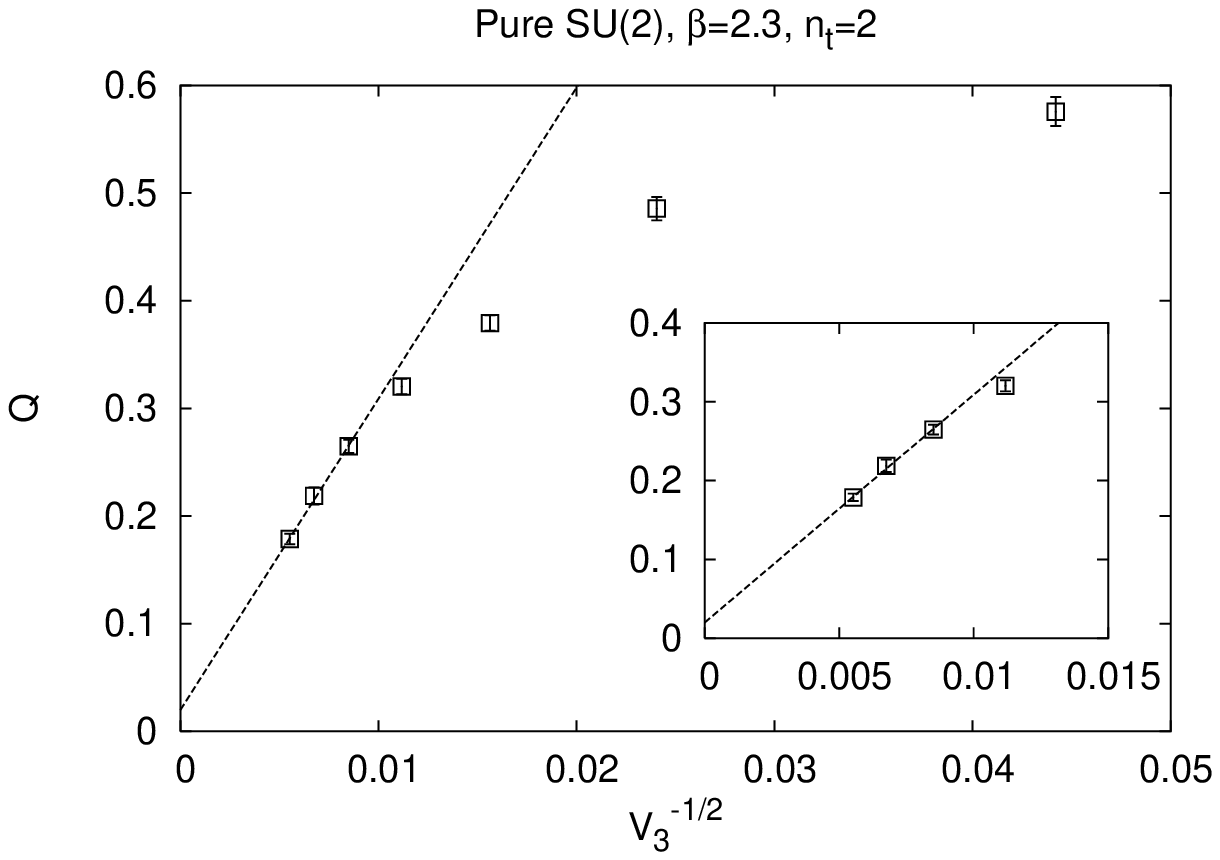}}}
\caption{The $Q$ parameter vs.\ root inverse 3-volume in the
high-temperature deconfined phase, pure SU(2) gauge theory at
$\b=2.3$ and $n_t=2$ lattice spacings.}
\label{q1}
}

\FIGURE[tb]{
\centerline{{\includegraphics[width=8truecm]{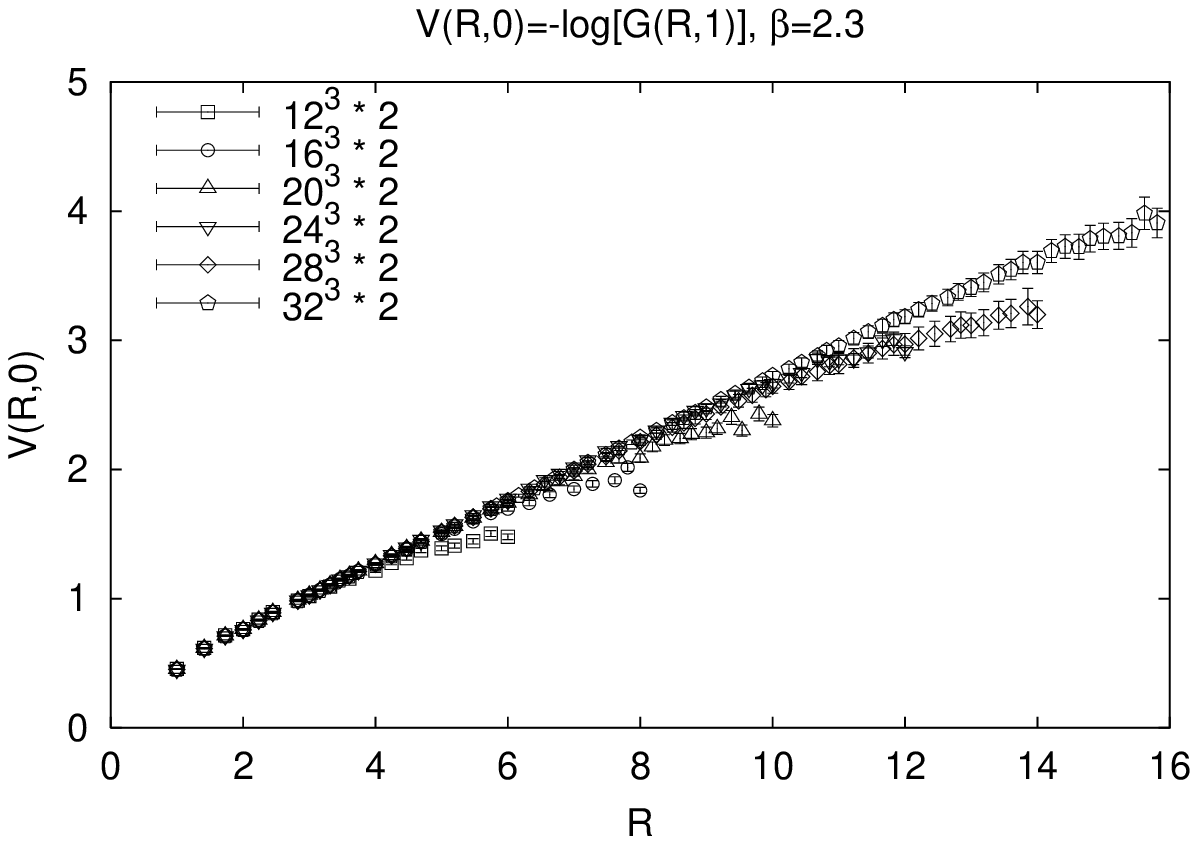}}}
\caption{$V(R,0)$ in the deconfined phase, at $\b=2.3$
with $n_t=2$ lattice spacings in the time direction, and space
volumes ranging from $12^3$ to $32^3$.}
\label{v1}
}

    The results for the Coulomb potential in the deconfined phase
are not paradoxical; we have already noted that remnant symmetry
breaking is a necessary but not sufficient condition for
confinement, and that the Coulomb potential is only an upper bound
on the static quark potential.  Thus it is possible for the
Coulomb potential to increase linearly even if the static quark
potential is screened, as evidently occurs in the deconfined
phase.  Nevertheless, this result is a little surprising, and it
would be nice to understand it a little better.

    Recall that in the continuum, the instantaneous part of the
timelike gluon propagator, which is proportional to the Coulomb
interaction energy between static charges, is given by eq.\ \rf{D}.
%
%
\ni Note that this is the expectation value of an operator which
depends only on the space components of the vector potential at a
fixed time.  On the lattice, this translates into an operator
which depends only on spacelike links on a time-slice.
However, we know that spacelike links, at fixed time, are a
confining ensemble, in the sense that spacelike Wilson loops have
an area law falloff even in the high-temperature deconfinement
phase.  If the confining property of the spacelike links on a
timeslice is not removed by the deconfinement transition, then it
is perhaps less surprising that the confining property of the
(latticized) operator in eq.\ \rf{D},
which depends only on spacelike links on a timeslice, survives in the
deconfinement regime.

    As a check, we apply a procedure that is known to remove the
confining properties of lattice configurations.  This is the de
Forcrand--D'Elia \cite{dFE} method of center vortex removal.  The
procedure is to first fix a given thermalized lattice
configuration to direct maximal center gauge, i.e.\ the gauge
which maximizes
\beq
       R = \sum_{x,\m} \left(\oh\mbox{Tr}[U_\m(x)]\right)^2
\eeq

\ni and carry out center projection

\beq
        Z_\m(x) = \mbox{signTr}[U_\m(x)]
\eeq

\ni to locate the vortices.  Vortices are then ``removed" from the
original configuration by setting

\beq
         U_\m(x) \ra U'_\m(x) = Z_\m(x) U_\m(x).
\eeq

\ni In effect this procedure superimposes a thin $Z_2$ vortex
inside the thick $SU(2)$ center vortices.  The effect of the thin
vortex is to cancel out the long range influence of the thick
vortex on Wilson loops.  It was found that this procedure not only
removes the Wilson loop area law falloff, but also removes
chiral symmetry breaking, and sends every configuration into the
zero topological charge sector \cite{dFE}.

    Having removed center vortices from the $L^3 \times 2$ lattice,
thereby removing the area-law falloff of spacelike Wilson loops,
we fix the modified configuration to Coulomb gauge, and compute
timelike link correlators in order to measure $V(R,0)$.  The
effect is quite dramatic.  It was found in ref.\ \cite{JS} that
vortex removal in pure gauge theory, in the low-temperature
confined phase, removes the confinement property of the Coulomb
potential. Now we see, from Fig.\ \ref{v01f}, that vortex removal also
removes the confining property of the Coulomb potential in the
high-temperature deconfinement phase. This is in accord with the
idea that it is the confining property of the ensemble of
spacelike links at fixed time (or, perhaps, the percolation of center
vortices on any time slice) that is crucial for the confining
property of the Coulomb energy.

\FIGURE[tb]{
\centerline{{\includegraphics[width=8truecm]{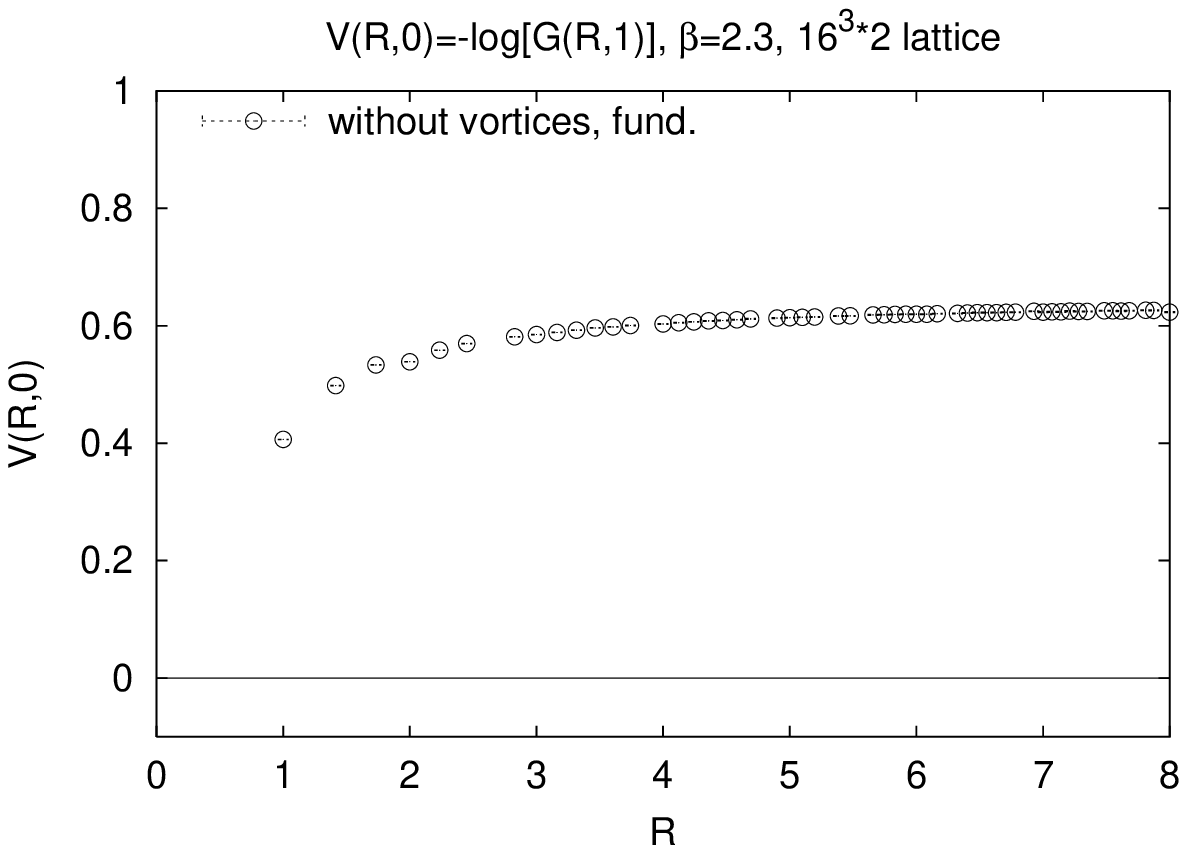}}}
\caption{The effect of vortex removal on $V(R,0)$
in the deconfined
phase, $\b=2.3$ on a $16^3 \times 2$ lattice.}
\label{v01f}
}

    It is interesting to ask whether there is some connection between the
center vortex confinement mechanism, and other proposals based on
the Gribov horizon \cite{horizon}.  While this question cannot yet be answered
definitively, there is one very intriguing fact that may be relevant: when
gauge-transformed
to the minimal Coulomb gauge, thin-vortex configurations lie
on the Gribov horizon, as we now explain.  \\

%
%
\section{Confinement Scenario in Coulomb Gauge and Vortex Dominance}
\label{confinement-scenario}

In the confined phase \cite{JS}, and in the deconfined
phase, Fig.~\ref{v1}, one sees clear evidence of a linearly rising color-Coulomb
potential.  There is a simple intuitive scenario in the minimal Coulomb
gauge that explains why $V_{coul}(R)$ is long range \cite{horizon}.  In
minimal Coulomb gauge the gauge-fixed configurations are (3-dimensionally)
transverse configurations that lie in the fundamental modular region $\L$.
In continuum gauge theory, $\L$ is convex and bounded in every direction
\cite{semenov}.  By simple entropy considerations, the population in a bounded
region of a high-dimensional space gets concentrated at the boundary.  For
example inside a sphere of radius $R$ in a $D$-dimensional space, the radial
density is given by $r^{D-1}dr$ and, for $r \leq R$, is highly concentrated
near the boundary $r = R$ for $D$ large.
With lattice discretization the dimension~$D$
of configuration space diverges like the volume $V$ of the lattice.
Moreover on the boundary the Faddeev--Popov
operator $M(A) = - \nabla \cdot D(A) $ has a vanishing eigenvalue,
as we will see.  (At large volume it has a
high density of small positive eigenvalues.)
This makes the color-Coulomb interaction kernel
$K(x, y; A) = M^{-1}(A)(- \nabla^2) M^{-1}(A)|_{xy}$ of long range for
typical configurations $A$ that dominate the functional integral.
We thus expect that
$V_{coul}(|\bx - \by|) = \langle K(x, y; A)\rangle $
is long range, although this
qualitative argument is not precise enough to establish that
$V_{coul}(R)$ rises linearly at large
$R$, as suggested by the numerical data.

We shall show that vortex dominance, which is strongly supported
by the data just presented, is consistent with this simple confinement scenario
in minimal Coulomb gauge.  More precisely we shall show that
{\it when a center configuration (defined below) is gauge transformed to
minimal Coulomb gauge it lies on the boundary $\pa \L$ of the fundamental
modular region $\L$.}  According to the confinement scenario in minimal Coulomb
gauge,
the probability measure is dominated by points at or near the
boundary $\pa \L$.
So center dominance, when translated into the minimal Coulomb gauge,
means dominance by a subset of configurations on the boundary $\pa \L$.
This is a stronger condition than the confinement
scenario in minimal Coulomb gauge, but consistent with it.

\paragraph*{Proof of assertion.}  To simplify the kinematics we give the continuum
version of the argument.  Numerical gauge fixing to minimal Coulomb gauge
corresponds to minimizing on each time slice the functional,
$F(A) = ||A||^2$, with respect to local gauge transformations.
Here  $||A||^2 = \int d^3x\;|A|^2$ is the square Hilbert norm of $A_i^a$.
At a minimum, which may be relative or absolute, the first variation
with respect to infinitesimal gauge transformations
$\d A_i = D_i(A) \omega$, vanishes for all $\omega$,
$\d ||A||^2 = 2(A_i, D_i(A) \omega) = 2(A_i, \pa_i \omega) = 0$,
which gives the Coulomb gauge condition $\pa_i A_i = 0$.  Moreover
at a relative or absolute minimum the second
variation with respect to gauge transformations
$\d^2 ||A||^2 = 2(D(A)_i \omega, \pa_i \omega) \geq 0$
is non-negative for all $\omega$, which is the statement that the
Faddeev--Popov operator $M(A) = - \nabla \cdot D(A)$ is non-negative.
These two conditions define the Gribov region $\Omega$,
\beq
\Omega \equiv \{A: \pa_i A_i = 0 \ {\rm and} \  - \pa_i D_i(A) \geq 0 \}.
\eeq
The fundamental modular region $\L$ is the set of absolute
minima with respect to gauge transformations,
\beq
\L \equiv \{A: ||A|| \leq ||{^g}A|| \ {\rm for \ all} \ g \}.
\eeq
It is included in the Gribov region,
$\L \subset \Omega$.  In the interior of $\Omega$
all eigenvalues of $M(A)$ are strictly positive $\l_n > 0$ (apart from the
trivial null eigenvalue associated with constant gauge transformations
$\pa_i \omega = 0$), and on the boundary $\pa \Omega$ there is a
non-trivial null eigenvector $\pa_i D_i(A) \omega_0 = 0$, and all other
eigenvalues are non-negative.\protect\footnote{The Gribov region $\Omega$ consists of
relative minima that are Gribov copies of the set $\Lambda$ of absolute minima.
Our numerical procedure selects the best Gribov copy obtained from eight random
gauge copies; there is no known
method for finding the absolute minimum.  On theoretical
grounds, one expects the sensitivity of our results to the choice of Gribov copy
be small \cite{Dan1}.  This expectation is susceptible to numerical investigation, as has
been done recently for the ghost propagator \cite{Bakeev}.}

We call a center configuration any lattice configuration $Z_{i}(x)$ for which
every link variable is a center element,
$Z_{i}(x) \in Z$ for every link $(x, \hat{\imath})$. The only non-zero action excitations of
center configurations are thin center vortices.
Such configurations are invariant
under all global gauge transformations $g^{-1}Z_{i}(x)g = Z_{i}(x)$.  Now apply
an arbitrary local gauge transformation $h(x)$ to the center configuration
$Z_{i}(x) \rightarrow V_{i}(x) = h^{-1}(x) Z_{i}(x) h({x + \hat{\imath}})$.  We
shall take $h(x)$ to be the gauge transformation that brings the center
configuration into the minimal Coulomb gauge.  In general, the transformed
configuration $V_{i}(x)$ is not an element of the center, but it is
invariant,
$V_{i}(x) = {g'}^{-1}(x) V_{i}(x) g'({x+\hat{\imath}})$,
with respect to the gauge transformation
$g'(x) = h^{-1}(x)gh(x)$ which, in general, is no longer global.

We give an infinitesimal characterization of the invariance of the
configuration $V_{i}(x)$ under the gauge transformations $g'(x)$.
The set of global gauge transformations form the SU($N$) Lie group and the
$g'(x) = \exp[\omega(x)]$ form a representation of this group.
Here $\omega(x)$ is an element of the Lie algebra of
SU($N$).  This algebra has $N^2 - 1$ linearly independent elements
$\omega^n(x)$, where $n = 1, \dots, N^2 -1$,
that satisfy $[\omega^l(x), \omega^m(x)] = f^{lmn}\omega^n(x)$.
Thus the configuration $V_{i}(x) = \exp[A_{i}(x)]$,
which is the gauge transform of the
center configuration $Z_{i}(x)$ into the minimal Coulomb gauge, is invariant
under local gauge transformations with $N^2 - 1$ independent
generators~$\omega^n$.\protect\footnote{The gauge orbit of the center
configuration is degenerate
and has $N^2-1$ fewer dimensions than a generic gauge orbit.}
This is the statement that in continuum notation
reads $A_i = A_i + \e D_i(A)\omega^n$, or $D_i(A) \omega^n = 0$.
It follows that the~$\omega^n$ also satisfy the weaker condition
$\nabla \cdot D(A) \omega^n = 0$.  Here $A$ is the representative in
minimal Coulomb gauge of the center configuration, and as such it lies
in the fundamental modular region $A \in \L$ (by definition) that moreover is
included in the Gribov region $\L \subset \Omega$, so we have
$A \in \Omega$.  However the equation $\nabla \cdot D(A) \omega^n = 0$
for $A \in \Omega$ means that~$A$ lies on its boundary~$\pa \Omega$.
With $\L \subset \Omega$ it follows that~$A$ lies on the
boundary~$\pa \L$ of~$\L$, at a point where the two boundaries touch.
We conclude that
the gauge transform of a center configuration lies on $\pa \L$, as
asserted.\protect\footnote{It should be noted that the numerical procedure
that we have used to remove center vortices is
4-dimensional.  The argument given here applies
in Coulomb gauge to center configurations
defined within 3-dimensional time-slices.}

The argument just given also applies to abelian configurations, namely a
configuration that lies in an abelian subalgebra of the Lie algebra.
Such a configuration
is invariant under a global U(1) gauge transformation.
{\it When an abelian configuration
is gauge-transformed to the minimal Coulomb gauge, it is mapped into
a point where the two boundaries $\pa \L$ and $\pa \Omega$ touch.}
But now the equation $D_i(A) \omega = 0$ has, in general, only one linearly
independent (non-trivial) solution, instead of $N^2 - 1$.  Thus abelian
dominance is also compatible with dominance of configurations on the
boundary $\pa \L$.

%
%
\section{SU(2) Gauge-Fundamental Higgs Theory}\label{gauge-fundamental-higgs}

    We now consider SU(2) gauge theory with the radially frozen Higgs field
in the fundamental representation.  For the SU(2) gauge group, the lattice
action can be written in the form \cite{Lang}
\bea
    S &=& \b \sum_{plaq} \oh \mbox{Tr}[UUU^\dg U^\dg] \non \\
      &+& \gamma \sum_{x,\m} \oh \mbox{Tr}[\phi^\dg(x) U_\m(x)
\phi(x+\widehat{\m})]
\eea
with $\phi$ an SU(2) group-valued field.  This theory cannot
be truly confining for non-zero $\gamma$, since the matter field
can screen any charge, and this simply reflects the absence of a
non-trivial global center symmetry. However, at sufficiently small
$\g$ there exists a ``pseudo-confinement" region, where the
static potential (as measured by the correlator of Polyakov loops)
is linear for some intermediate range of quark
separations before the onset of screening. At large $\gamma$
there is a Higgs region, where the linear potential is completely
absent. It was shown many years ago by Fradkin and Shenker
\cite{FS} that any two points in the Higgs and pseudo-confinement
regions can be joined by a path in the $\b-\gamma$ coupling plane
that avoids all thermodynamic singularities.  Although there exists a
line of first-order phase transitions in the $\b-\g$ plane,
this line has an endpoint and does not divide the diagram into
thermodynamically separate phases.  (It appears as the solid line in
Fig.~\ref{phase_f}, below.)  In accord with the
Fradkin-Shenker observation, numerical simulations suggest that
there is only one non-confining phase in the gauge-fundamental
Higgs theory.

     In apparent contradiction to this fact, one can make a strong
case for the existence
of a remnant symmetry-breaking transition at small $\b$.
We set $\b = 0$ so the action is
\beq
S_\phi = \gamma \sum_{x,\m} \oh
\mbox{Tr}[\phi^\dg(x) U_\m(x) \phi(x+\widehat{\m})].
\eeq
We shall show that (i) at large $\g$ there is spontaneous breaking
of the remnant gauge symmetry, associated with the short range of $V_{coul}(R)$,
but (ii) at small $\g$ there is a linear rise of $V_{coul}(R)$ at large $R$.
Thus along the line $\b = 0$ in the $\b-\g$ plane
we expect a transition at some finite value of $\g$.

(i) First consider the limit
$\gamma \rightarrow \infty$.  The action $S_\phi$ is a maximum when
$U_\m(x) = \phi(x) \phi^\dg(x + \widehat{\m})$
holds on each link, and
when $\g$ gets large, $U_\m(x)$ gets frozen at this value.  This configuration
is a gauge-transform of the identity $U_\m(x) = I$, and when fixed to the
minimal Coulomb gauge the spatial components get fixed to the identity
$U_i(\bx, t) = \phi(\bx, t) \phi^\dg(\bx + \hat{\imath}, t) = I$,
for all $\bx$ and $i$.  Thus $\phi(\bx, t)$ is independent of~$\bx$.
We write $\phi(\bx, t) = g(t)$, and we have
$U_0(\bx, t) = g(t)g^\dg(t+1)$, which is also independent of~$\bx$.
This is the gauge analog of all spins aligned, and we expect
spontaneous breaking of the remnant gauge symmetry.  Indeed, we have
\beq
\tilde{U}(t) = \frac{1}{V_3}\sum_{\bx} U_0(\bx, t) = g(t)g^\dg(t+1),
\eeq
and the order
parameter $Q$ defined above has the value $Q = 1$.  This is maximal
breaking of the remnant gauge symmetry.

(ii)  Now consider small values of $\g$.  We shall calculate the lattice analog
of $V_{coul}(R)$, namely $V(R, 0) = - \log[G(R, 1)]$,
to leading non-zero order in $\g$.  The gauge fixing involves only
the spatial link variables $U_i$, and with the action $S_\phi$ the
integration over the $U_0$ factorizes into a product over link integrals.
To evaluate
$G(R, 1) = \langle\oh {\rm Tr}[U_0(\bx, 1) U_0^\dg(\by, 1)]\rangle$,
we first integrate over the $U_0(\bx, t)$,
with the result, to leading order in $\g$,
\beq
G(R, 1) =  \frac{\g^2}{16}\left\langle \oh\mbox{Tr}
\left[\phi^\dg(\bx, 2)\phi(\bx, 1) \phi^\dg(\by, 2)\phi(\by, 1)\right] \right\rangle.
\eeq
There are now 4 unsaturated $\phi$ fields.  For simplicity we suppose
that $(\bx, 1)$ and $(\by, 1)$ are joined by a principle axis,
which we take to be the 1-axis, and $R = |\bx - \by|$.  The
leading contribution to the $\phi$-integration at small $\g$ is obtained by
saturating each link on the line that runs from $(\bx, 1)$ to
$(\by, 1)$ by ``bringing down" from the exponent~$S_\phi$ the term
$\oh \g\; \mbox{Tr}[\phi({\bf z}, 1) U_1({\bf z}, 1)
\phi^\dg({\bf z} + \widehat{1}, 1)]$,
and likewise for the line from $(\bx, 2)$ to $(\by, 2)$.  This gives a factor
$\g^{2R}$.  The $\phi$-integrations are now effected.
The remaining integration on the
spatial link variables $U_i$ is finite because of the gauge fixing.  We
cannot evaluate it explicitly, but this last integration does not introduce
any further $\g$-dependence.  We thus obtain $G(R, 1) = \g^{2R + 2} \times
H(R)$.  Here $H(R)$ is not known, but it is independent of~$\g$.  This gives
$V(R,0) = - (2R+2)\log\g - \log[H(R)]$.  The asymptotic fall-off in the
correlator $G(R, 1)$ should not be more rapid than exponential, so $\lim_{R
\rightarrow \infty} \log[H(R)] = -c R$.  We thus obtain a linear rise at
large $R$ in $V(R, 0) \sim \s_{coul} R$, where the ``Coulomb" string
tension is given by
\beq
\s_{coul} = - 2 \log \g + c.
\eeq
This is
non-zero for small~$\g$.  To make this argument rigorous one would have to
show that the expansion of $G(R, 1)$ in powers of $\g$ converges.  However
this calculation does strongly suggest that at $\b = 0$ the remnant
symmetry is unbroken for small $\g$ whereas, as we have seen, it is broken
at large~$\g$.%
\protect\footnote{One can make a similar calculation of $V(R, 0)$ for small $\b$
(strong coupling) in pure SU($N$) gauge theory.  This yields a finite
``Coulomb" string tension at small $\b$ given by
$\s_{coul} = - \log \b$.
This suggests that at least in the strong-coupling region
there is a ``Coulomb flux tube" that connects the external sources.}

Returning to the SU(2) gauge-fundamental Higgs theory, we note that the
above calculations at large and small $\g$ are easily understood in terms of
the Coulomb-gauge and center-dominance scenarios.  Indeed,
for large $\g$ (and any $\b$), the gauge-fixed configurations are at
or near $U_i(x) = I$.  This configuration is an interior point of the
fundamental modular region~$\L$.  Thus at large~$\g$ the coupling to the
fundamental Higgs is effective in keeping configurations away from the
boundary~$\pa \L$, where the thin vortex configurations are to be found, and
where the Faddeev--Popov operator is of long range.  On the contrary, at
small $\g$ the coupling to the Higgs field is ineffective, and
entropy leads to dominance
of configurations on the boundary~$\pa \L$.

    Now we turn to numerical simulation.
Figure \ref{q2p1} is a calculation of $Q$ vs.\ $\g$ at $\b=2.1$,
where it is known (from the work of Lang et al.\ \cite{Lang}) that
the first-order transition is around $\g=0.9$.  Below $\g=0.9$,
$Q$ seems to extrapolate to zero at infinite volume, while above
the transition $Q$ extrapolates to a non-zero constant. There
appears to be an actual discontinuity in $Q$ across the whole line
of first-order (thermodynamic) phase transitions in the $\b - \g$
coupling plane.  At sufficiently small values of $\b$, there is no
thermodynamic transition, and we see no discontinuity in $Q$ as a
function of $\g$.  What we see instead is that $Q\approx 0$ over a
finite range of $\g$, and then, beyond a critical value
$\g=\g_{cr}$, $Q$ smoothly increases with increasing $\g$.  Our
results for $Q$ vs.\ $\g$ at $\b=0$, on $8^4$ and $16^4$
lattices, are shown in Fig.\ \ref{qbeta0}; the solid line is the
presumed extrapolation to infinite volume. If we were dealing with
a spin system, and $Q$ were the magnetization, this would clearly
represent a second order phase transition.  In the present case it
is certainly a symmetry-breaking transition, separating a
symmetric region with $Q=0$ from a broken-symmetry region of
$Q>0$.

\FIGURE[tb]{
\centerline{{\includegraphics[width=8truecm]{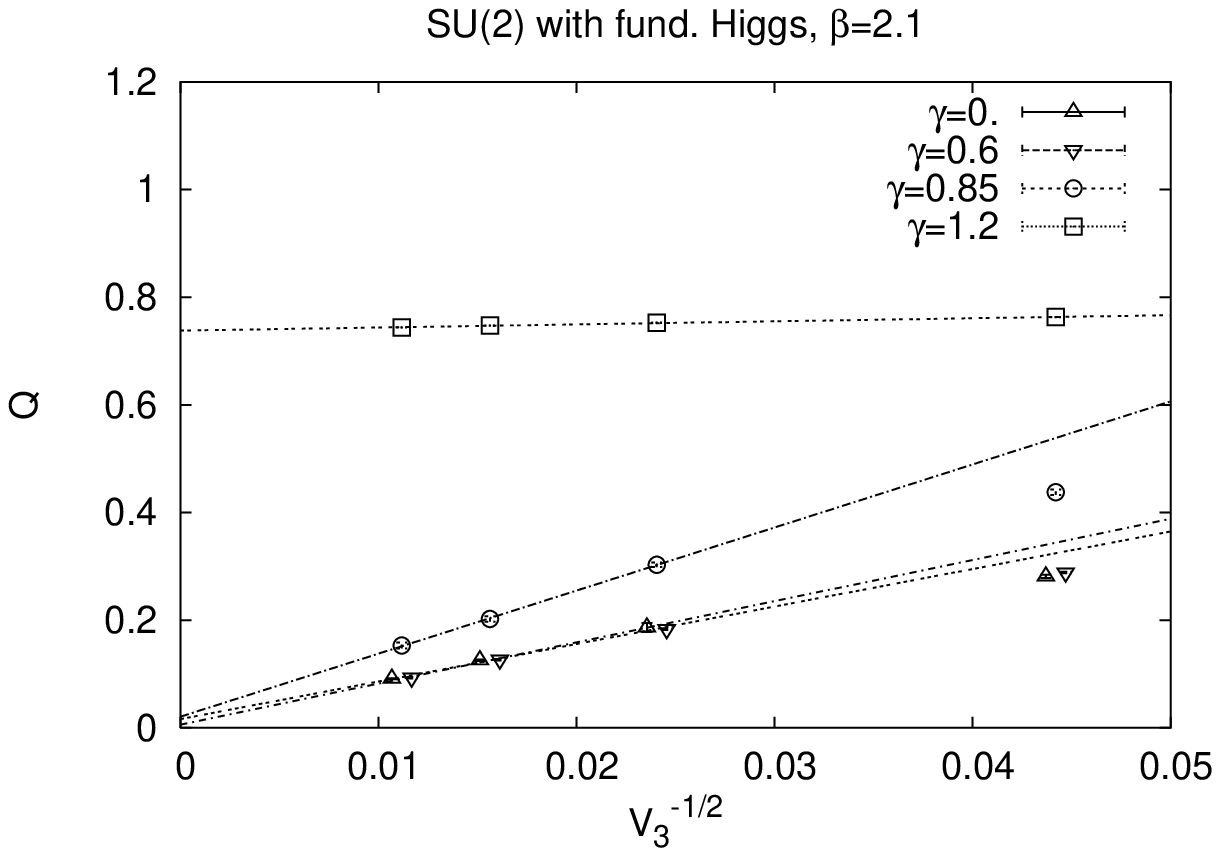}}}
\caption{Plot of $Q$ vs.\ root inverse 3-volume, and
extrapolation of $Q$ to infinite volume in the
gauge-fundamental Higgs theory at $\b=2.1$ and various $\gamma$,
above ($\g=1.2$) and below the first-order transition point
around $\g=0.9$.}
\label{q2p1}
}

\FIGURE[tb]{
\centerline{{\includegraphics[width=8truecm]{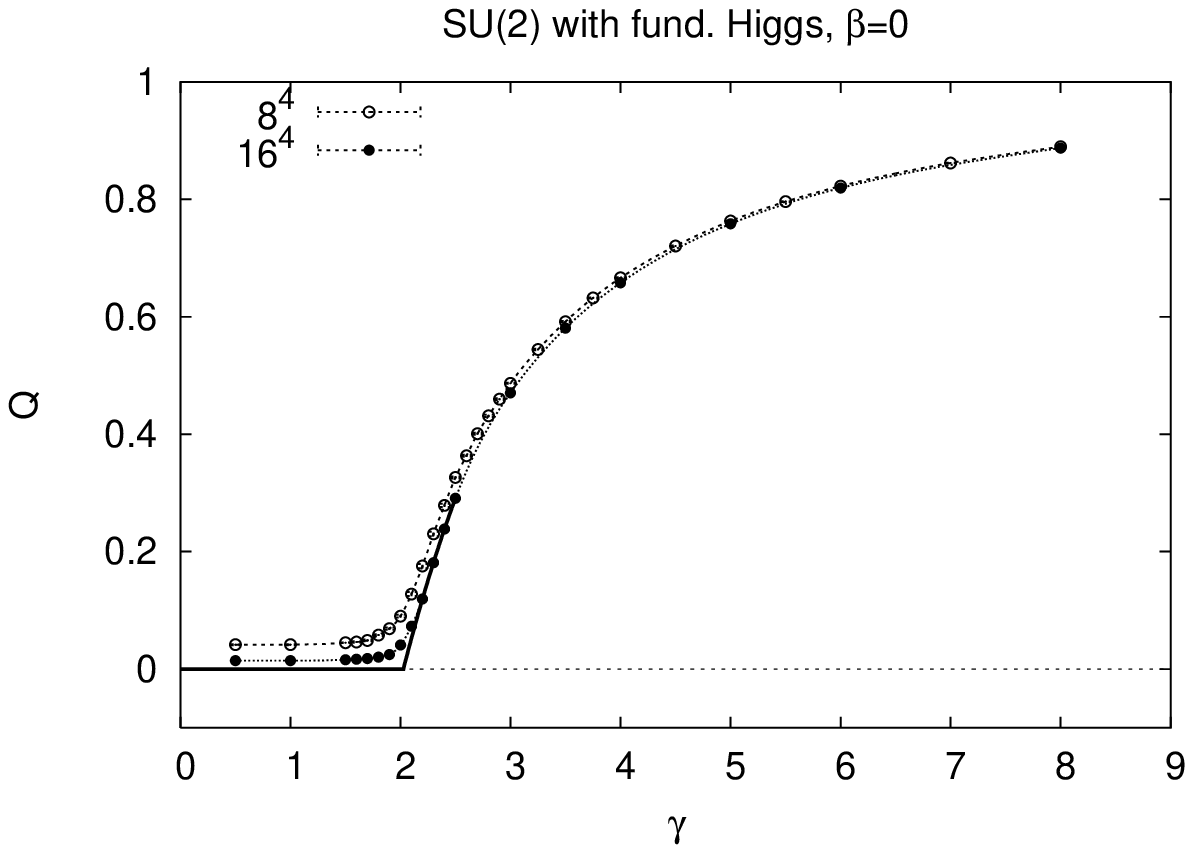}}}
\caption{$Q$ vs.\ $\g$ at $\b=0$ in the SU(2) fundamental Higgs
model, on $8^4$ and $16^4$ lattices.  The solid line is the
presumed extrapolation of $Q$ to infinite volume.}
\label{qbeta0}
}

   On the other hand, despite the existence of a symmetry-breaking
transition, there is no thermodynamic transition of any kind at
$\b=0$.  At this value of $\b$ the free energy can be computed
exactly, with the result, in a 4-volume $V$
\beq
           F(\g) = 4 V \log\left[ 2I_1(\g) \over \g\right]
\eeq
which is perfectly analytic for all $\g > 0$.  Thus we have
confirmed the theoretical argument that there must be a remnant
symmetry-breaking transition even at small $\b$, but we have also
found that this transition is not accompanied (at small $\b$) by a
thermodynamic transition, defined as some degree of
non-analyticity in the free energy.  Our result for the line of
critical couplings of the remnant symmetry-breaking transition is
shown in Fig.\ \ref{phase_f}.  Along the solid line there is also
a thermodynamic (first-order) transition, which is absent along the
dashed line.

\FIGURE[tb]{
\centerline{{\includegraphics[width=8truecm]{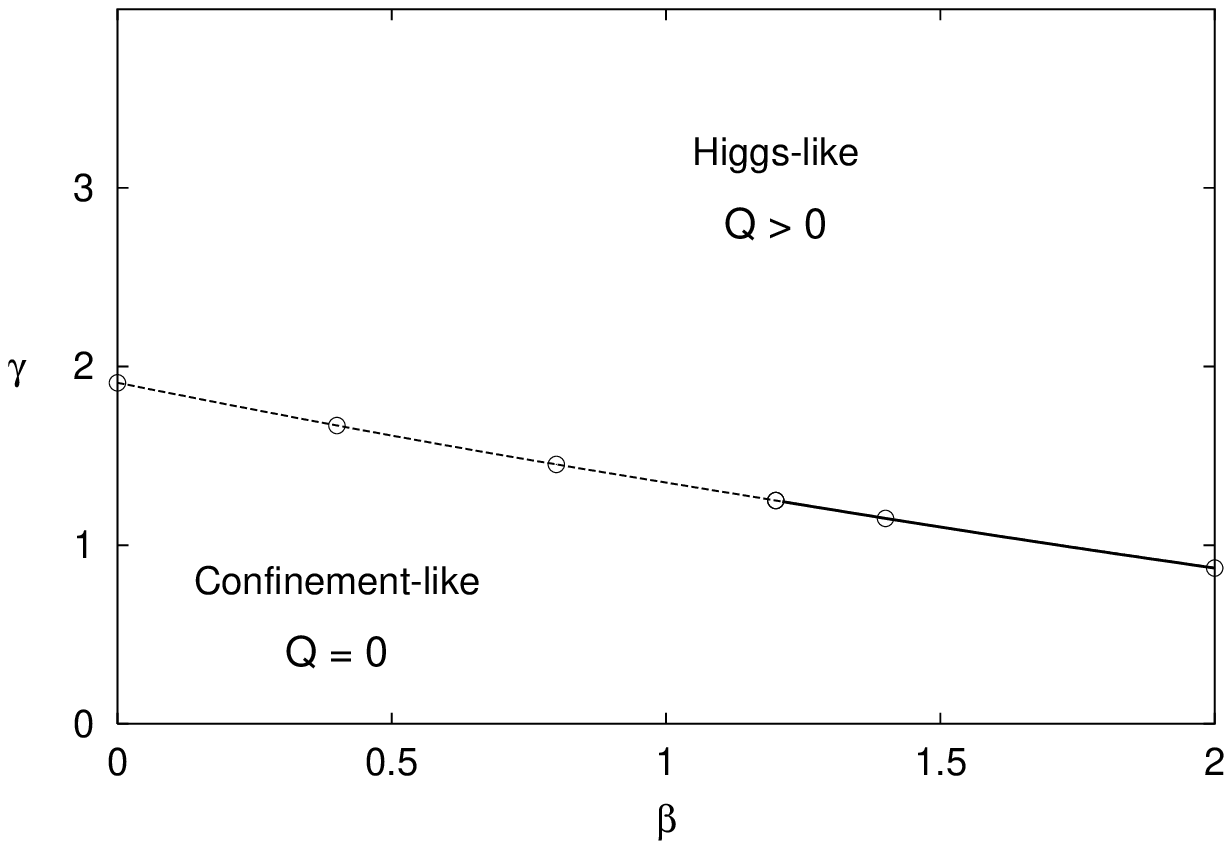}}}
\caption{Phase diagram of the SU(2) fundamental Higgs model.
There is a thermodynamic transition and a $Q$ (remnant
symmetry-breaking) transition along the solid line, but only
a non-thermodynamic transition
(Kert\'esz line) along the dashed line.}
\label{phase_f}
}

     Our result for the gauge fundamental Higgs system is not entirely
new; it was in fact anticipated by Langfeld in ref.\ \cite{Kurt1},
who considered a closely related model in Landau (rather than
Coulomb) gauge. In that work the modulus of the space-time
averaged Higgs field was used as an order parameter to detect the
breaking of remnant symmetry, which in Landau gauge must be both
space \emph{and} time independent.  In ref.\ \cite{Kurt2} it was
further suggested that the line of remnant symmetry breaking
transitions, where it is unaccompanied by a line of thermodynamic
transitions, is a Kert\'esz line \cite{Kertesz}. A Kert\'esz line
is a line of percolation transitions;  the original example comes
from the Ising model.  In the Ising model, in the absence of an
external magnetic field, there is a phase transition from a $Z_2$
symmetric to an ordered phase, and this transition can be
expressed, in different variables, as a transition from a
percolating phase at low temperature, to a non-percolating phase
at high temperature.  In the presence of a magnetic field, the
partition function and thermodynamic observables become analytic
in temperature; there is no thermodynamic phase transition.
Nevertheless, the percolation transition persists, and traces out
a Kert\'esz line in the temperature-magnetic field plane,
completely separating the phase diagram into two regions.  But if there
is a Kert\'esz line in the
gauge-Higgs coupling plane, the question is what kind of
objects are percolating.
Based in part on results reported by Bertle and Faber
\cite{Bertle1}, Langfeld \cite{Kurt2} proposed that the unbroken
remnant symmetry region is a region of percolating center
vortices, which cease percolating in the broken symmetry region.
There is now very good evidence for a vortex percolation
transition of this kind in gauge-fundamental Higgs theory,
reported in ref.~\cite{Bertle2}.\protect\footnote{The role of Kert\'esz
lines in high-temperature QCD is also discussed by Satz in ref.\
\cite{Satz}.}

     Our findings here support the idea that there is some physical
distinction that can be made between the Higgs and the
pseudo-confining regions of the gauge-fundamental Higgs phase
diagram. In the pseudo-confining region the remnant symmetry is
unbroken, the Coulomb potential rises linearly, and center
vortices percolate, while the Higgs region is a region of broken
symmetry, the Coulomb potential is asymptotically flat, and center
vortices do not percolate. This distinction appears to exist
despite the fact that the two regions are thermodynamically
connected, as demonstrated by Fradkin and Shenker in
ref.\ \cite{FS}.

     Before leaving the gauge-fundamental Higgs theory, we should ask
what are the effects of vortex removal in the symmetric phase. We
already know that the Coulomb potential in the fundamental Higgs
theory must be confining in the symmetric (pseudo-confinement)
phase and screened in the broken phase; the only issue is how the
Coulomb potential is affected in each phase if center vortices are
removed. In Figs.\ \ref{v1a} and \ref{v1b} we see the Coulomb
potential in the symmetric ($\b=2.1, ~ \g=0.6$) and broken
($\b=2.1,~\g=1.2$) phases, respectively, before and after vortex
removal. In the symmetric phase, vortex removal by the de
Forcrand--D'Elia procedure sends the Coulomb string tension to
zero, as in the high-temperature phase of pure gauge theory.  Deep
in the Higgs phase, on the other hand, the effect of vortex
removal is seen to be very minor.

\FIGURE[tb]{
\centerline{{\includegraphics[width=8truecm]{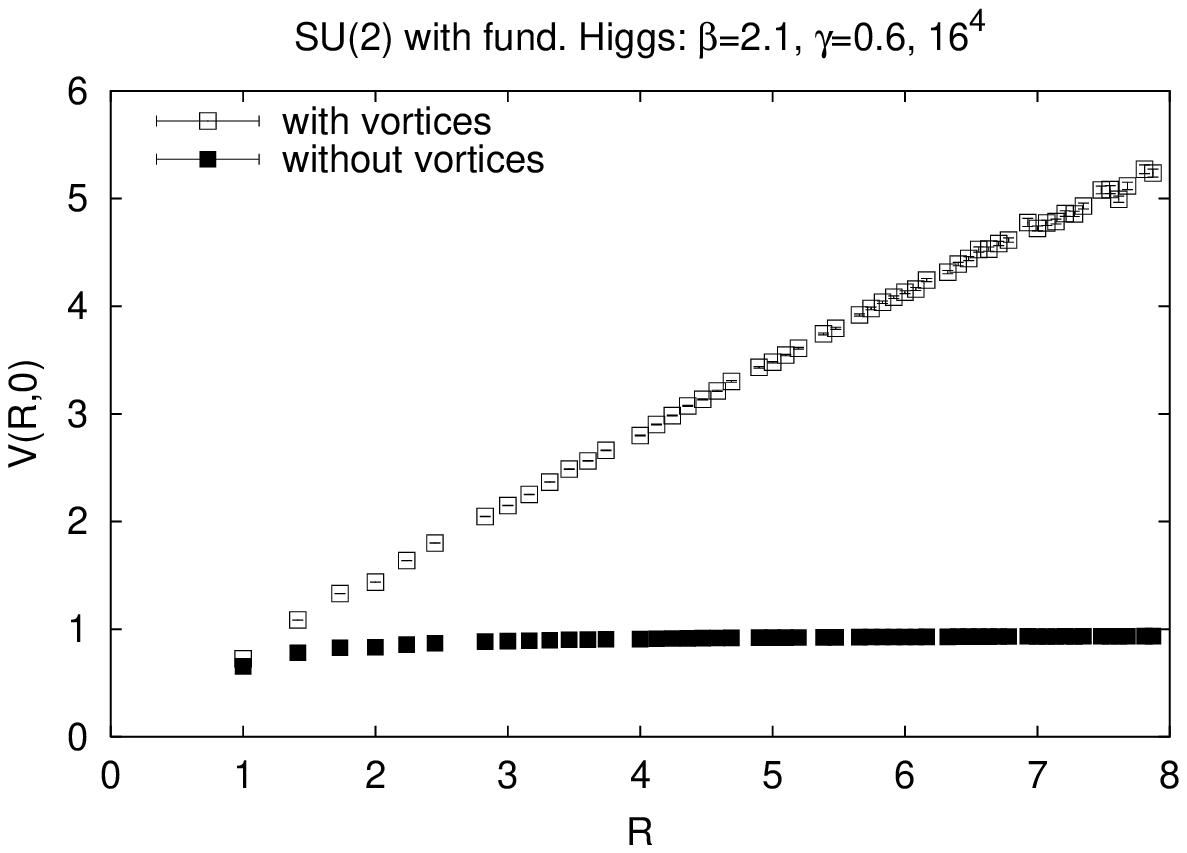}}}
\caption{Gauge-fundamental Higgs theory: effect of vortex removal
in the symmetric (pseudo-confinement) phase, $\b=2.1,~\g=0.6$.}
\label{v1a}
}

\FIGURE[tb]{
\centerline{{\includegraphics[width=8truecm]{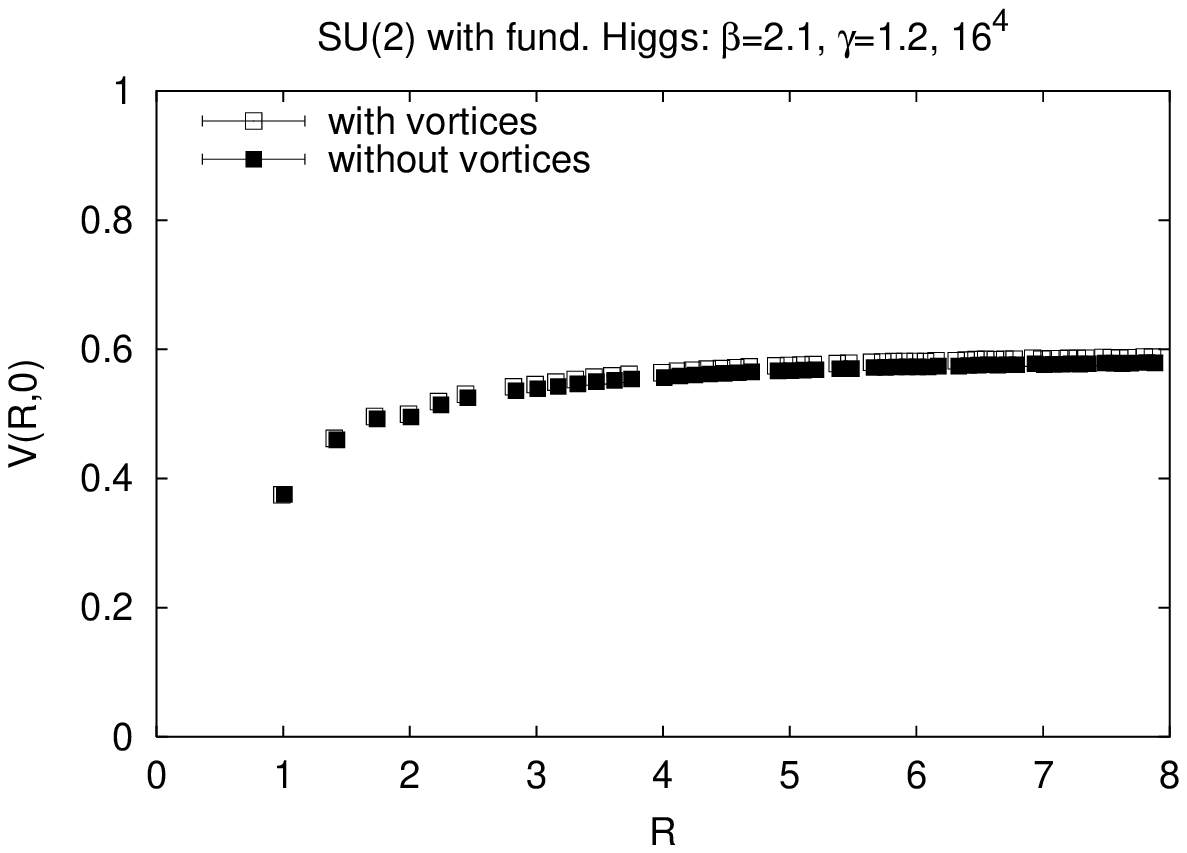}}}
\caption{Gauge-fundamental Higgs theory: effect of vortex removal
in the broken (Higgs) phase, $\b=2.1,~\g=1.2$.}
\label{v1b}
}

%
%
\section{The Adjoint-Higgs Model Revisited}\label{gauge-adjoint-higgs-2}

    We have understood theoretically why there should be a remnant
symmetry-breaking transition at small $\b$ in the gauge-fundamental
Higgs model, as seen in the numerical data.  This
raises immediately the question of why there is no corresponding
transition in $Q$, at small $\b$, in the gauge-adjoint Higgs
theory.  Now in the fundamental-Higgs model, the transition at
small $\b$ appears to be a percolation transition, as discussed in
the previous section. But the absence of a percolation transition
at low $\b$ in the adjoint-Higgs case is easy to understand. There
the Higgs part of the action is insensitive to the existence
of thin center vortices, and cannot suppress their condensation at
small $\b$ regardless of the value of $\g$. More concretely, at
$\b=0$, the action is invariant with respect to local
transformations $U_0(x) \ra z(x) U_0(x)$ where $z(x)=\pm 1$, and
this immediately implies that $Q=0$ at infinite volume, again
regardless of the value of $\g$.

    On the other hand, if the absence of a remnant
symmetry-breaking transition at low $\b$ is due to large fluctuations of
center elements $z(x)$, then one might still expect breaking, at low
$\b$ and large $\g$, of the SO(3)=SU(2)/$Z_2$ part of the remnant
symmetry group, which is insensitive to $U_0(x) \ra z(x) U_0(x)$
fluctuations.  The relevant order parameter is
\beq
     Q_{adj} = {1\over n_t} \sum_{t=1}^{n_t} \left\langle
     \sqrt{\oth \mbox{Tr}\left[\tU_{adj}(t)
     \tU_{adj}^\dg(t)\right]} \right\rangle,
\eeq
where $\tU_{adj}(t)$ is the spatial average of timelike links
in the adjoint representation
\bea
        \tU_{adj}(t) &=& {1\over V_3} \sum_{\bx} U_{0,adj}(\bx,t),
\non \\
     U^{ab}_{0,adj}(\bx,t) &=& \oh \mbox{Tr}
\left[\s^a U_0(\bx,t) \s^b U_0^\dg(\bx,t)\right].
\eea

    Rather surprisingly, there appears to be no transition in $Q_{adj}$
either, in the adjoint-Higgs model.  As we see in Fig.\ \ref{qadj8},
for data taken
on a small $8^4$ lattice, there is no sign of any transition for $Q_{adj}$ at
finite $\g$ and $\b=0$.  In fact, $Q_{adj}$ is essentially $\g$-independent.
Extrapolation of $Q_{adj}$ to infinite volume anywhere in the confined ($Q=0$)
phase is consistent with $Q_{adj}=0$, as seen in Fig.\ \ref{qadjvsL}. Thus
the transition line for $Q_{adj}$ in the $\b - \g$
coupling plane appears to be the same as for $Q$ in Fig.\ \ref{phase_a}, but
not like $Q$ in Fig.\ \ref{phase_f}, and this fact calls for an
explanation.

\FIGURE[tb]{
\centerline{{\includegraphics[width=8truecm]{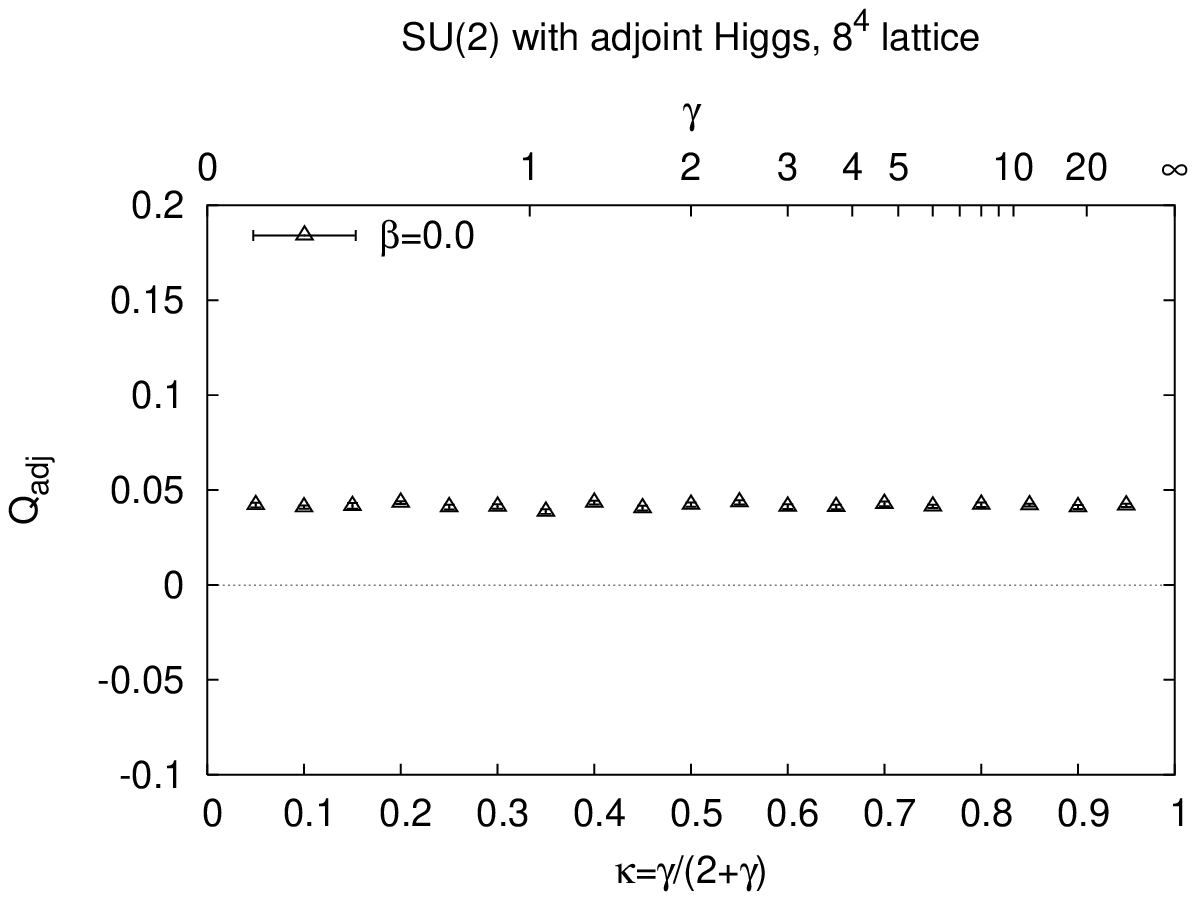}}}
\caption{(Lack of) variation of $Q_{adj}$ with $\g$ at $\b=0$
in the gauge-adjoint Higgs theory.}
\label{qadj8}
}

\FIGURE[tb]{
\centerline{{\includegraphics[width=8truecm]{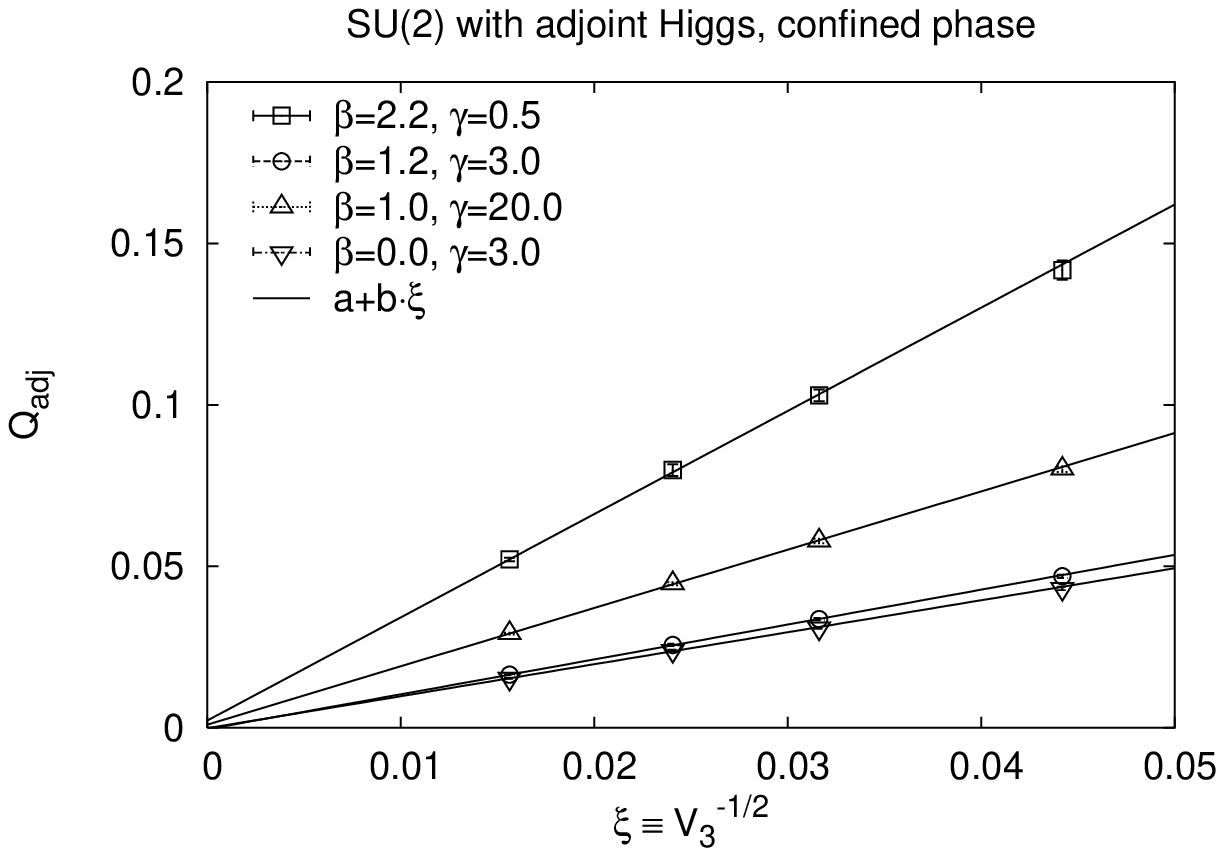}}}
\caption{Plot of $Q_{adj}$ vs.\ root inverse 3-volume, and
extrapolation of $Q_{adj}$ to infinite volume, at a variety
of couplings in the confined phase, in gauge-adjoint Higgs theory.}
\label{qadjvsL}
}

The difference in the phase diagrams of Figs.~3 and 12 occurs at
small~$\b$ and large~$\g$.  To understand this difference
we set $\b = 0$ in the adjoint Higgs action \rf{adjhiggs} so it reads
\beq
S_\phi = \frac{\g}{2} \sum_{x,\m} \phi^a(x) U_{\m, adj}^{ab} \phi^b(x + \widehat{\m}),
\eeq
and we shall evaluate the equal time correlator
in minimal Coulomb gauge
\beq
G_{adj}(\bx - \by, 1) = \left\langle\oth
\mbox{Tr}\left[U_{0,adj}(\bx, 1) U_{0, adj}^\dg(\by, 1)\right]\right\rangle
\eeq
for $\g$ large.

\paragraph*{Step 1.  Integration over $U_0$.}
The gauge fixing does not involve the $U_0$, so the
integrals over the $U_0$ factorize, with the result on each
time-like link
\bea
  \langle U_{0,adj}^{ab}(x) \rangle & = &
      \l \ \langle \phi^a(x) \phi^b(x + \widehat{0}) \rangle,   \non  \\
   \l & = &\frac{\cosh (\frac{\g}{2})}{\sinh (\frac{\g}{2})} - \frac{2}{\g} \ .
\eea
Here $\l$ has the limiting value $\l = 1$ at large $\g$. The
correlator factorizes into a product of expectation values
on adjacent time slices
\bea
&&G_{adj}(\bx - \by, 1) = \oth\left\langle
\phi^a(\bx, 1) \phi^b(\bx, 2)
\phi^b(\by, 2) \phi^a(\by, 1) \right\rangle     \non   \\
      &&= \oth\; \Bigl\langle\phi^a(\bx, 1) \phi^a(\by, 1)\Bigr\rangle\;\left\langle\phi^b(\bx, 2) \phi^b(\by, 2)\right\rangle.
\eea
This gives
\beq
G_{adj}(\bx - \by, 1)  =  3 C^2(\bx - \by),
\eeq
where
\beq
C(\bx - \by) = \oth\left\langle\phi^a(\bx, 1) \phi^a(\by, 1) \right\rangle
\eeq
is the $\phi$-propagator within a single time slice.  It is calculated
using the action within a time slice,
\beq
S_{slice} = \frac{\g}{2} \sum_{\bx,i}
\phi^a(\bx) U_{i,adj}^{ab}(\bx) \phi^b(\bx + \hat{\imath}).
\eeq
(The argument $t$ that enumerates time slices is suppressed in the following.)

\paragraph*{Step 2.  Reduction to U(1) theory.}  To evaluate the $\phi$-propagator
at large $\g$, we introduce a gauge transformation $g(\bx) \in$ SU(2) that
depends on $\phi(\bx)$ with the property that the group element in the adjoint
representation, denoted $\R^{ab}[g(\bx)]$, is the SO(3)
rotation  matrix that rotates the 3-direction into the $\phi$-direction,
$\phi^a(\bx) = \R^{a3}[g(\bx)]$.
The action and correlator  are given by
\bea
S_{slice} & = & \frac{\g}{2} \sum_{\bx,i}
\R^{33}\left[g^{-1}(\bx) U_i(\bx) g(\bx + \hat{\imath})\right]  \non \\
C(\bx - \by) & = & \oth\left\langle
  \R^{a3}[g(\bx)] \R^{a3}[g(\by)]  \right\rangle.
\eea
These expressions are invariant under $g(\bx) \rightarrow g(\bx)h(\bx)$,
where $h(\bx)$ is an SU(2) element that corresponds to a rotation about the
3-axis.  So we may average over $h(\bx)$ which results in replacing
the integral over $\phi(\bx)$ by an integral
over $g(\bx) \in$ SU(2).  At large $\g$ the functional integral is
dominated by the maximum of the action $S_{slice}$, which occurs where
$\R^{33}[g^{-1}(\bx) U_i(\bx) g(\bx + \hat{\imath})] = 1$
holds on each link, namely, where
$u_i(\bx) \equiv g^{-1}(\bx) U_i(\bx) g(\bx + \hat{\imath})$
is a rotation about the 3-axis.  These rotations form the U(1) group.
Thus at the maximum of the action
the link variables are given by
\bea
U_i(\bx) & = & g(\bx) u_i(\bx) g^{-1}(\bx + \hat{\imath}) \non \\
& = &{^g}u_i(\bx),
\eea
where $g(\bx) \in$ SU(2).  Thus $U_i(\bx)$
is an SU(2) gauge transform of $u_i(\bx) \in$ U(1).
At large $\g$ the $U_i(x)$ get frozen into this form,
and the integral over $U_i(\bx) \in $ SU(2) gets reduced to an
integral over abelian configurations $u_i(\bx) \in $ U(1).
We have noted that we may replace $g(\bx)$ by
$g(\bx)h(\bx)$, where $h(\bx) \in$ U(1).  We use this freedom to gauge
fix the $u_i(\bx)$ within the U(1) group of rotations about the 3-axis.
Naturally we choose
the minimal Coulomb gauge of U(1) gauge theory.   Thus the integral over
the $U_i(\bx) \in$ SU(2) gets replaced by an integral over
gauge-fixed configurations $u_i(\bx) \in $ U(1).  We designate this set by $T$.
In continuum gauge theory, $T$ is the set of {\it all} transverse
abelian configurations $A_i^3(\bx)$.  In sharp contrast to the SU(2) case, in
abelian gauge theory there are no Gribov copies.
Different transverse configurations $A_i^3(\bx)$ are gauge-inequivalent,
and the set $T$ of gauge-fixed abelian configurations
$A_i^3(\bx)$ is the unbounded set of {\it all} transverse
abelian configurations, without restriction to a fundamental modular region.
(This is a major difference between
abelian and non-abelian gauge theories, and is the basis of the confinement
scenario in minimal Coulomb gauge.)

\paragraph*{Step 3.  Integration over the $g(\bx)$.}
Although~$u \in T$ is completely gauge-fixed by minimizing with respect to local $U(1)$
gauge transformations, it is not gauge-fixed by minimizing
with respect to local SU(2) gauge transformations, because that is a larger
group.  But our calculation in minimal Coulomb gauge requires that
the SU(2) configuration $U_i(x) = {^g}u_i(\bx) \in \L$ be completely
gauge-fixed with respect to local SU(2) gauge transformations.  So
for given~$u$, ~$g(\bx)$ is the unique SU(2) gauge transformation
(modulo global gauge transformations) that accomplishes this gauge fixing.
We write $g(\bx) = g(\bx; u)$, and the $\phi$-propagator is given by
\beq
C(\bx - \by) = \int_T du \ P(u)  \
\oth \R^{a3}[g(\bx; u)] \R^{a3}[g(\by; u)],
\eeq
where $P(u)$ is a positive probability density.

\paragraph*{Step 4.  Integration over the $u(\bx)$.}
We saw in section~\ref{confinement-scenario}
that when configurations in an abelian subgroup
are gauge-fixed, they get mapped into the boundary $\pa \L$.
However that conclusion is an over-statement which ignores the
fact that some~U(1) configurations~$u \in T$ lie in the interior of~$\L$
(although this happens with probability 0, as we shall see).
Since in the continuum limit~$\L$ is a subset of transverse SU(2)
configurations $A_i^{a, tr}(\bx)$
that is bounded in every direction
whereas $T$ is the unbounded set of {\it all} transverse abelian configurations
$A_i^{3, tr}(\bx)$, it follows that
some configurations $u \in T$ lie
inside $\Lambda$ and some lie outside $\Lambda$.
Correspondingly we break up
the integral into contributions from $u \in T$ inside $\L$, and from
$u \in T$ outside $\L$,
\bea
C(\bx - \by) & = & C_{in}(\bx - \by) + C_{out}(\bx - \by)  \non \\
C_{in}(\bx - \by) & = & \int_{T \cap \L} du \ P(u)  \
    \oth \R^{a3}[g(\bx; u)] \R^{a3}[g(\by; u)]  \non \\
C_{out}(\bx - \by) & = & \int_{T - T \cap \L} du \ P(u)\times\non \\
    &&\oth \R^{a3}[g(\bx; u)] \R^{a3}[g(\by; u)].
\eea
For $u \in T \cap \L$ the unique gauge transformation $g(\bx,u)$ that brings
$u$ inside $\L$ is the identity transformation
\beq
g(\bx, u) = I; \ \ \ \  u \in T \cap \L.
\eeq
This gives $\R^{a3}[g(\bx; u)] = \R^{a3}[I] = \d_{a3}$, and we
obtain
\bea
C_{in}(\bx - \by) & = & \int_{u \in T \cap \L} du \ P(u)  \non \\
     & = & P_{in},
\eea
where $P_{in}$ is that probability that $u \in T$ lies inside $\L$.
This is independent of~$\bx$ and~$\by$, and corresponds to ordered spins.
It resembles the calculation with coupling to fundamental Higgs
at $\b = 0$ and $\g$ large, where $P_{in} = 1$.
For $u$ outside~$\L$ it appears that the solution~$g(\bx, u)$ of the
spin-glass minimization problem depends in a very irregular and disordered
way on $u$ and $\bx$, so $C_{out}(R)$ decays rapidly,
$\lim_{R \rightarrow \infty}C_{out}(R) = 0$.  With
$G_{adj}(R,1) = 3 C^2(R)$ this gives
\beq
\lim_{R \rightarrow \infty} G_{adj}(R,1) = 3 P_{in}^2.
\eeq
Thus, remarkably, a numerical determination of $G_{adj}(R,1)$
provides a direct measurement of the probability $P_{in}$ that a configuration
$u \in T$ lies inside $\L$.

The data of Fig.~\ref{qadjvsL} strongly suggest that $Q_{adj}$
extrapolates to 0 at infinite volume.  This is the
disordered phase, in which
$\lim_{R \rightarrow \infty} G_{adj}(R,1) = 0$ holds.  This gives
\beq
\lim_{V \rightarrow \infty} P_{in} = 0.
\eeq
We have noted that in continuum gauge theory~$\L$ is bounded in every
direction whereas~$T$ is unbounded in all
directions.  According to the simple entropy estimate,
$p(r)dr \sim r^{D-1}dr$, where $D$ is the (very high) dimension of
the space~$T$ of transverse~U(1) configurations,
the fraction~$P_{in}$ of the set~$T$ that lies inside
the bounded region~$\L$ is negligible, $P_{in} \ra 0$, as
in fact the data indicate, and we have also $P_{out} \ra 1$.
Moreover an abelian configuration
$u \in T$ that lies outside~$\L$, gets gauge-transformed
in the minimal Coulomb gauge into a
configuration $U = {^g}u$ that lies on the
boundary~$\pa \L$, as was shown in section~\ref{confinement-scenario},
so in this instance all the probability lies on
the boundary $\pa \L$ and the measure of the interior vanishes.
This exemplifies the simple scenario
in Coulomb gauge, according to which confinement occurs when
the functional integral is dominated by the boundary~$\pa \L$.

The absence at~$\b = 0$ of a transition in~$Q_{adj}$
as~$\g$ increases from~0 to~$\infty$ is now explained.  For
with coupling to the adjoint Higgs, the measure of ordered spins
that would break the remnant gauge symmetry is~$P_{in} = 0$
at large $\g$, and the remnant gauge symmetry is preserved.
By contrast, with coupling to the fundamental Higgs for $\g$ large,
$U_i(x)$ is a gauge transform of the identity $I$, as we have seen in
section~\ref{gauge-fundamental-higgs}.  Since $I$ is certainly in $\L$,
then $P_{in} = 1$, and the remnant symmetry is maximally broken. The coupling
to the fundamental Higgs field at large~$\g$ keeps configurations
away from the boundary $\pa \L$.

   Finally we wish to emphasize that with coupling to the adjoint Higgs, the numerical
result $P_{in}=0$ is a direct manifestation of the deep difference between the
fundamental modular regions (defined by minimizing $F[A]$ with respect to local gauge
transformations) of an abelian and non-abelian gauge theory.  The fundamental
modular region is unbounded in every direction in an abelian gauge theory, but
bounded in every direction in a non-abelian gauge theory.

%
%
\section{Compact QED}\label{qed}

   Although in the preceding sections we have considered only the
theories with a non-abelian SU(2) gauge invariance, there is no
barrier to studying remnant symmetry breaking in an abelian model
such as compact $QED_4$.  In this
model we have confinement at strong couplings and a massless phase at
weak couplings, with a transition between the two phases at
approximately $\b=1$.  In Fig.\ \ref{qv} we show our results for (the abelian analogue of)
$Q$ at $\b=0.7$, which is inside the confinement phase, and at
$\b=1.3$, which lies in the massless phase.  The results are as
expected; $Q$ extrapolates nicely to zero at infinite volume in
the confined phase, and appears to extrapolate to a non-zero value
in the massless phase.  Figure \ref{prop07} is a plot of $V(R,0)$
vs.\ charge separation $R$ at $\b=0.7$ on variety of hypercubic
lattice volumes $L^4$. Note that the potential is insensitive to
changes in lattice volume, and that deviations from the linear potential,
where they are statistically significant, arise from the lack of
rotation invariance at strong couplings.

\FIGURE[tb]{
\centerline{{\includegraphics[width=8truecm]{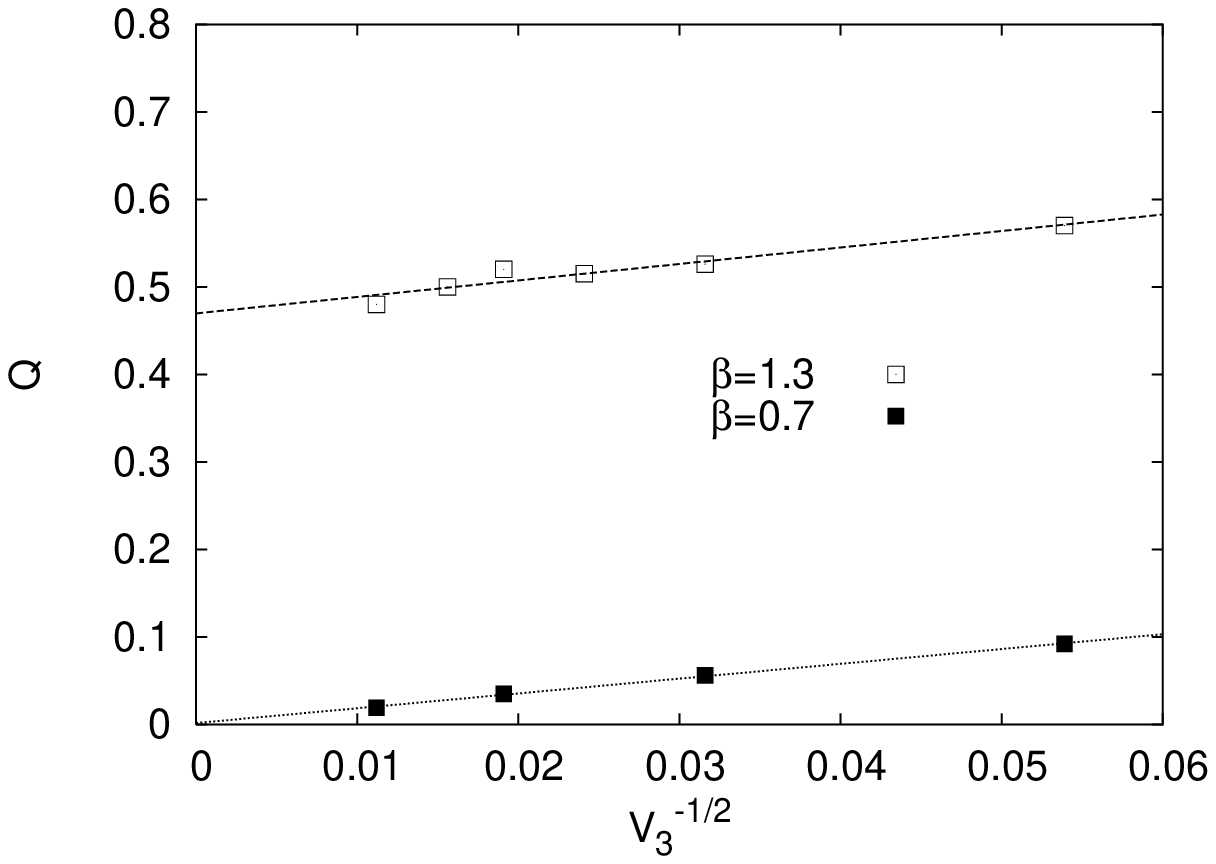}}}
\caption{Plot of $Q$ vs.\ root inverse 3-volume, and
extrapolation of $Q$ to infinite volume in $QED_4$,
for $\b=0.7$ (confining phase) and $\b=1.3$ (massless phase).}
\label{qv}
}

\FIGURE[tb]{
\centerline{{\includegraphics[width=8truecm]{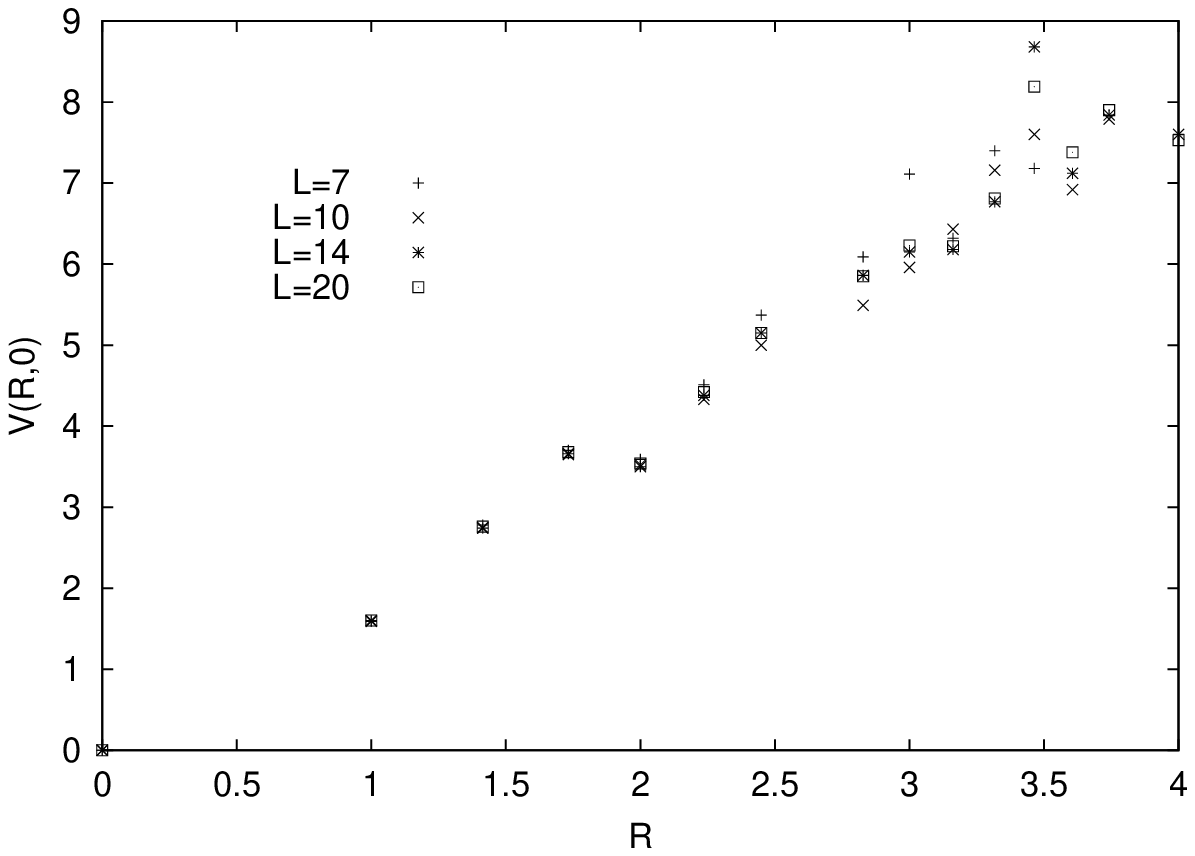}}}
\caption{The potential $V(R,0)$ in $QED_4$ at $\b=0.7$, on
a variety of $L^4$ lattice volumes.}
\label{prop07}
}

\FIGURE[tb]{
\centerline{{\includegraphics[width=8truecm]{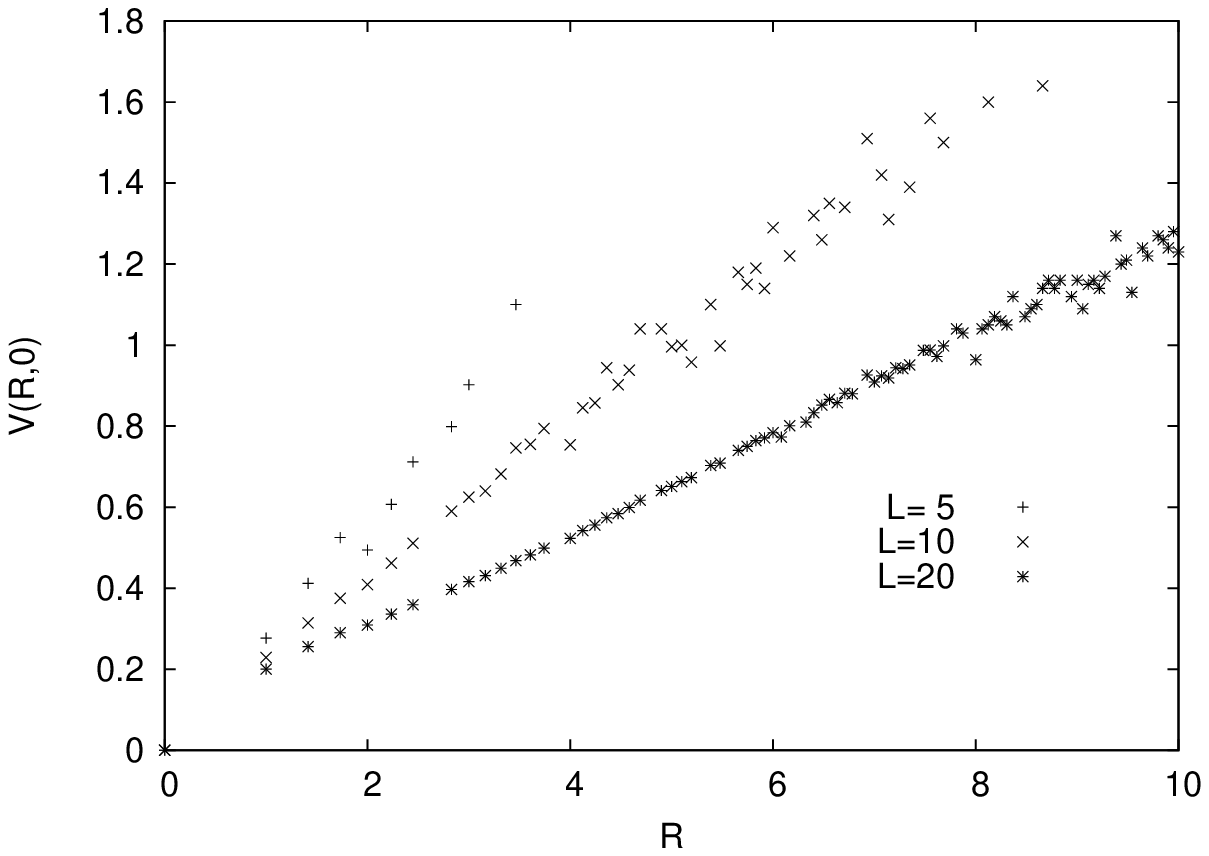}}}
\caption{The potential $V(R,0)$ in $QED_4$ at $\b=1.3$, on
several $L^4$ lattice volumes.}
\label{prop13} }

    The situation changes drastically in the massless phase at $\b=1.3$,
where our results for $V(R,0)$ are displayed for $5^4,~10^4$ and
$20^4$ lattices in Fig.\ \ref{prop13}. Although the string tension
extracted from $V(R,0)$ appears to be non-zero on each of these
lattices (which cannot be correct for the Coulomb potential in the
weak-coupling regime), this string tension drops very markedly (by
about 1/3) with each doubling of the lattice volume. This volume
dependence of the string tension is very different from what we
have observed for the non-abelian theories in various phases.  For
the abelian theory, the
expected result $\s_{coul} = 0$ is presumably recovered
in the large volume limit.  In fact, from the value of
$Q$ extrapolated to infinite volume, we can estimate that on a
large lattice $V(R,0)$ should be bounded, very roughly, by
$-\ln(Q^2) \approx 1.6$ at large $R$. This bound implies that
$V(R,0)$ has to level out in large volumes, resulting in
$\s_{coul}=0$.

    What we learn from the weak-coupling case is
that $V(R,0)$ may have very significant
lattice size dependence in theories such as $QED_4$
in the massless phase, where the correlation length is large (comparable
to lattice size) or infinite.
It is therefore important to compute $V(R,0)$ on a variety of
lattice sizes, and to extrapolate $Q$ to infinite volumes, as we have
done in the preceding sections.

%
%
\section{Translation to Temporal Gauge and Stringless States}\label{translation}

In this section we shall show how to translate back and
forth between the minimal Coulomb gauge and the temporal gauge, $A_0 = 0$,
so the measurements reported here have an equivalent description in the
temporal gauge.  Both of these gauges are compatible with a Hamiltonian
formulation and a
physical transfer matrix.  Moreover the temporal gauge is invariant under
all space-dependent but time-independent
gauge transformations~$g(\bx)$, where~$\bx$ is a 3-vector, and it may be helpful
to express things in a more gauge-invariant way.  We shall see that
the state obtained here by numerically
gauge-fixing to the minimal Coulomb gauge
becomes, after translation into temporal gauge, a ``stringless" state
of the type introduced by Lavelle and McMullan \cite{lavelle}
that does not involve a Wilson line joining the sources.

\subsection{Temporal gauge}\label{temporal-gauge}

In temporal gauge the continuum Hamiltonian has the
canonical form
\beq
H = \frac{1}{2g^2} \int d^3x \ (E^2 + B^2).
\eeq
Here the spatial components $A_i^a(x)$ of the vector potential are
canonically conjugate to the color-electric field operator
$E_i^a(x) = i \frac{\d}{\d A_i^a(x)}$, and
$B_i^a = \e_{ijk}(\pa_j A_k^a + \oh f^{abc}A_j^b A_k^c)$.

The color-charge density of the static quarks
does not appear in $H$ because $A_0 = 0$.  Instead it appears in
the generator of local space-dependent gauge transformations,
\beq
G^a(x) = - D^{ac}(A) \cdot E^c(x) + \r_q^a(x),
\eeq
where
\beq
\r_q^a(x) = t_1^a \d(x-x_1) + t_2^a \d(x-x_2)
\eeq
is the color-charge density of a pair of external quarks at~$x_1$ and~$x_2$.
The wave functional $\Psi_{\a\b}(A)$ bears the color indices of the
external quarks, on which the quark charge vectors act according to
\beq
  (t_1^a\Psi)_{\a\b} = t_{\a\g}^a \Psi_{\g\b}; \qquad
(t_2^a\Psi)_{\a\b} = -t_{\b\g}^{a*} \Psi_{\a\g},
\eeq
where the $t^a$ form the basis of an irreducible representation of the Lie
algebra of SU($N$), $[t^a, t^b] = if^{abc}t^c$.  One may verify that
$G(\omega) = \int d^3x \ \omega^a(x) G^a(x)$ generates an infinitesimal gauge
transformation,
\bea
\left[G(\omega), A_i^a\right] &=& i D_i^{ac} \omega^c; \non \\
\left[G(\omega), E_i^a \right] &=& i f^{abc} E_i^b\omega^c;  \non \\
\left[G(\omega), \r_q^a\right] &=& i f^{abc} \r_q^b\omega^c.
\eea
This transformation leaves $H$ invariant, $[G(\omega), H] = 0$,
so we may choose wave-functionals that transform irreducibly under the
local gauge group.  Physical wave-functionals are required to satisfy
the subsidiary condition
\beq
G(x) \Psi = 0,
\eeq
which is both Gauss' law and the statement that the wave-functional is
gauge-invariant.  This condition determines the gauge-transformation
properties of the wave-functional
\beq
\Psi({^g}A) = g(x_1) \Psi(A)g^\dg(x_2),
\label{transform}
\eeq
where we use matrix notation for the quark color indices, and
${^g}A_i = g A_i g^\dg + g \pa_i g^\dg$.

\subsection{From temporal gauge to minimal Coulomb gauge, and back}\label{temporal-coulomb}

The continuum temporal gauge does not really provide a well-defined
quantum mechanics because the inner product for gauge-invariant
wave functionals
\beq
(\Psi_1, \Psi_2) = \int dA \ \Psi_1^*(A) \Psi_2(A)
\eeq
diverges due to the local gauge invariance of the wave-functionals.
Gauge-fixing is required to correctly normalize the wave-functionals in
temporal gauge.  We do this
by applying the Faddeev--Popov formula to gauge-invariant inner products.
For this purpose we parametrize configurations by
$A = {^g}A^{\rm tr}$, where $A^{\rm tr}$ is the
representative of $A$ in the minimal Coulomb gauge, so
$A^{\rm tr} \in \L$ is a transverse configuration in the fundamental
modular region.  The Faddeev--Popov formula gives
\beq
(\Psi_1, \Psi_2) = \int_\L dA^{\rm tr} \ \det M(A^{\rm tr})
\ \Psi_1^*(A^{\rm tr}) \Psi_2(A^{\rm tr}).
\eeq
Here $M(A^{\rm tr}) = - \nabla \cdot D(A^{\rm tr})$
is the Faddeev--Popov operator, which is symmetric and positive for
$A^{\rm tr} \in \L$.  The right hand side is the inner product in minimal
Coulomb gauge.  Thus, in the 3-dimensional operator formalism, the minimal
Coulomb gauge is a gauge-fixing {\it within} the temporal gauge of the
3-dimensional local gauge invariance,
and {\it the wave functional in minimal
Coulomb gauge is the restriction of the wave-functional in temporal gauge
to the fundamental modular region}\protect\footnote{To avoid confusion we
note that the two conditions $A_0 = 0$ and $\pa_i A_i = 0$
can be imposed at a {\it fixed} time, which is sufficient for the
3-dimensional operator formalism.  They cannot both be maintained for
{\it all} time, and in the 4-dimensional Feynman path-integral formalism
in Coulomb gauge one maintains $\pa_i A_i = 0$, but $A_0 \neq 0$.}
\beq
\Psi_{coul}(A^{\rm tr}) = \Psi(A^{\rm tr});  \ \ \ \ A^{\rm tr} \in \L.
\eeq

Conversely, {\it the gauge invariance of the wave-functional in
temporal gauge defines the unique extension of
the wave-functional in minimal Coulomb gauge
into a wave-functional in temporal gauge.}
We parametrize an arbitrary configuration by
$A = {^g}A^{\rm tr}$, where $A^{\rm tr} \in \L$, and
$g(x; A)$ and $A^{\rm tr}(x; A)$ depend on the
configuration $A$.  The existence and uniqueness of these quantities
at the non-perturbative level is assured
(with lattice regularization) by the existence of an absolute
minimum with respect to gauge transformations of
$F_A(g) = ||{^{g^{-1}}}A||^2$.
 From \rf{transform} above we obtain,
in matrix notation,
\bea
\Psi(A) & = &  \Psi[{^g} A^{\rm tr}] \non   \\
& = &  g(x_1; A) \ \Psi[A^{\rm tr}(A)] \ g^\dg(x_2; A),
\label{temporal}
\eea
which expresses the wave-functional in temporal gauge in terms of the
wave-functional $\Psi(A^{\rm tr})$
in minimal Coulomb gauge.  The gauge transformation
$g(x; A)$ that is found numerically when gauge-fixing to the minimal
Coulomb gauge has reappeared in the wave-functional in the temporal gauge.

[For completeness, we note that a
quick way to obtain the Hamiltonian in Coulomb gauge \cite{Christ}
is to apply the Faddeev--Popov formula to the matrix elements of~$E^2$,
\beq
(E_i^a \Psi_1, E_i^a \Psi_2) =
\int_\L dA^{\rm tr} \ \det M(A^{\rm tr})
(E_i\Psi_1)^* E_i\Psi_2|_{A = A^{\rm tr}}.
\label{electricenergy}
\eeq
 From this formula the matrix elements of the Hamiltonian in Coulomb gauge
are easily found once $E_i\Psi|_{A = A^{\rm tr}}$ is specified.
To evaluate $E_i\Psi|_{A = A^{\rm tr}}$ we solve Gauss' law for
the longitudinal part of the color-electric field.  We write
$E_i = E_i^{\rm tr} - \pa_i \ph$, where $\ph^a(x)$ is the color-Coulomb
potential operator, and
$E_i^{\rm tr} = i \frac{\d}{\d A^{\rm tr}}$ satisfies
\beq
[ E_i^{a,{\rm tr}}(x), A_j^{b,{\rm tr}}(y) ]
=i [\d_{ij} - \pa_i \pa_j (\pa^2)^{-1}] \ \d(x-y) \ \d^{ab}.
\eeq
Gauss' law in temporal gauge reads
\beq
- D_i \pa_i\ph \Psi + A_i \times E_i^{\rm tr}\Psi = \r_q \Psi,
\eeq
where $(X \times Y)^a = f^{abc}X^bY^c$ is the Lie bracket.
We solve for the color-Coulomb potential
\beq
\ph(x) \ \Psi|_{A = A^{\rm tr}}
  =  [M^{-1}(A^{\rm tr}) \r](x) \ \Psi(A^{\rm tr}),
\eeq
where $A^{\rm tr} \in \L$.
Here $M(A^{\rm tr}) = - D(A^{\rm tr}) \cdot \pa
= - \pa \cdot D(A^{\rm tr})$
is the Faddeev--Popov operator, and
\beq
\r \equiv  - A_i^{\rm tr} \times E_i^{\rm tr} + \r_q
\eeq
is the total color-charge density of quarks and dynamical gluons.
This gives the desired expression,
\beq
E_i  \ \Psi|_{A = A^{\rm tr}} =
  [E_i^{\rm tr} - \pa_i M^{-1}(A^{\rm tr}) \r] \ \Psi(A^{\rm tr}),
\eeq
which is to be used in \rf{electricenergy}.]

\subsection{Energy in Coulomb gauge is energy in temporal gauge}\label{correspondence}
The quantity we have measured is the expectation value

\beq
\langle H_{coul} \rangle - E_0 =  V_{coul}(|\bx-\by|) + E_{se}
\eeq
in the state with wave-functional in minimal Coulomb gauge
\beq
\Psi_{\a\b}(A^{\rm tr}) = {\textstyle{\frac{1}{\sqrt{2}}}} \ \d_{\a\b} \ \Phi_0(A^{\rm tr}),
\eeq
where $\Phi_0(A^{\rm tr})$ is the vacuum state of pure glue.

We translate this back into temporal gauge.
 From \rf{temporal} we obtain
\beq
\Psi_{\a\b}(A) = {\textstyle{\frac{1}{\sqrt{2}}}} \ [g(x_1; A) g^\dg(x_2; A)]_{\a\b} \  \Phi_0(A),
\label{stringless}
\eeq
where we have used the gauge invariance of the vacuum state of pure glue,
$\Phi_0(A^{\rm tr}) = \Phi_0({^g}A^{\rm tr}) = \Phi_0(A)$.  Thus the
quantity we measure may equivalently
be described as the expectation-value of the Hamiltonian in temporal gauge,
with this wave functional
\beq
(\Psi, H \Psi) - E_0 = V_{coul}(|\bx - \by|) + E_{se}.
\eeq

\subsection{Stringless states}\label{stringless-states}

A wave-functional of type \rf{stringless},
with $\Phi_0(A)$ an arbitrary gauge-invariant
scalar function, has been considered before \cite{lavelle} and was
called a ``stringless" state.\protect\footnote{It was called
``stringless" because in its construction the thin string,
$P\exp(\int_{x_1}^{x_2} A_i dx^i)$,
was replaced by $g(x_1; A) g^\dg(x_2; A)$ which transforms in the same way
under gauge transformation.  It was argued
that the thin string has infinite energy, whereas the ``stringless"
state has finite energy (after ultraviolet renormalization) and is a
better approximation to the correct hadron state.
However when $A$ lies on a degenerate orbit,
${^h}A = A$ for some $h(x) \neq I$,
the parametrization $A = {^g}A^{\rm tr}$ is singular.
As a result the stringless wave-functional is singular for such
configurations, which may raise its energy significantly.}
The motivation for constructing this state was that it has the correct
gauge-transformation properties, which is equivalent to Gauss' law being
satisfied exactly.  Moreover in QED,
with external charges only, the stringless state is the exact
wave-functional, with
$g(x; A) = \exp[i (\nabla^2)^{-1} \nabla \cdot A]$.
This leads us to expect that in QCD the stringless state becomes exact at
short distance $|\bx-\by|$.  Thus the stringless state has several attractive
properties.

The stringless state was originally constructed using a
perturbative expansion for~$g(x; A)$.  This expansion does not converge
for~$A^{\rm tr}$
outside the fundamental modular region~$\L$.  However, as we have noted,
the existence
of $g(x; A)$ at the non-perturbative level is assured  by the minimization
procedure, and $g(x; A)$ is the gauge transformation
that we have found numerically.

Since our numerical data strongly
suggest that $V_{coul}(R)$ rises
linearly, it appears from \rf{stringless}
that the ``stringless" state of quarks manifests a
finite string tension $\s_{coul}$
at large separation.  Our numerical finding is
that this string tension exceeds the standard string tension $\s$
by $\s_{coul} \sim 3 \s$.  This provides a measure of the extent to which
the stringless state~\rf{stringless} fails to be the exact ground state
of a pair of external quarks.

%
%
\section{Conclusions}\label{conclusions}

    In this article we have shown that the confining property
of the color-Coulomb potential is tied to the unbroken realization
of a remnant global gauge symmetry in Coulomb gauge. We have
studied this type of confinement numerically in SU(2) gauge-Higgs
theories, and in pure gauge theory at zero and at finite temperatures.
Confinement in the color-Coulomb potential is not identical to
confinement in the static quark potential.  We have seen that the
deconfined phase in pure gauge theory, and the pseudo-confinement
region of gauge-fundamental Higgs theory, are instances in which
the color-Coulomb potential is asymptotically linear, even though
the static quark potential is screened. In terms of symmetries, the
point is that center symmetry breaking, spontaneous or explicit, does not
necessarily imply remnant symmetry breaking.

     The existence of a confining color-Coulomb potential, in
cases where the static quark potential is screened, has some
bearing on the question: In what sense does confinement exist in
real QCD, with dynamical quarks?  The problem is that in gauge
theories with matter fields
in the fundamental representation, such as real QCD, there is no
non-trivial center symmetry, and no possibility of having an
asymptotically confining static potential. Further, in
gauge theories with a scalar matter field in the fundamental
representation, there is no local order
parameter that can distinguish between the Higgs and confinement
phases, and the Fradkin-Shenker theorem assures us that there is
no thermodynamic transition of any kind that can isolate the
Higgs phase from a distinct confinement phase.\protect\footnote{While it
is sometimes suggested that confinement should simply be
understood as the condition that asymptotic particle states are
all color singlets, this condition is also fulfilled in the Higgs
phase of gauge-Higgs theories (cf., e.g., ref.\ \cite{review}).}
All this suggests that there is no fundamental difference between
real QCD and SU(3) gauge theory in a Higgs phase, and it should be
possible to interpolate smoothly from the one theory to the other
by a continuous change of parameters.

    The results reported in this paper suggest otherwise.  In gauge
theories with matter fields in the fundamental representation, the
``confinement" phase and the Higgs phase are distinguished by the
symmetric or broken realization of a remnant gauge symmetry.
Remnant symmetry breaking is not accompanied by non-analyticity in
the free energy, nevertheless the (non-local) order parameter $Q$
is directly related to the color-Coulomb potential, which is
confining in the $Q=0$ phase. Thus, in real QCD, the gluon propagator in
Coulomb gauge is confining. This confining property of (dressed)
one-gluon exchange is absent in the Higgs phase of a gauge field
theory.

    We have also uncovered a strong correlation between the presence
of center vortices and the existence of a confining Coulomb
potential.  Thin center vortices, as pointed out above, lie on the
Gribov horizon. In every case where we find a confining Coulomb
potential, we also find that removal of center vortices takes the
Coulomb string tension to zero. In related work on the SU(2)
gauge-fundamental
Higgs theory, it was suggested \cite{Kurt2}, and recently verified
\cite{Bertle2}, that vortices
percolate throughout the lattice in the pseudo-confined
phase, and do not percolate in the Higgs phase. These phases
correspond to regions of unbroken and broken remnant symmetry,
which are completely separated by a percolation transition
located along a Kert\'esz line.

    There are many open questions.  Presumably the linear color-Coulomb
potential is associated with a flux tube of longitudinal color
electric field.  If this is really so, does the tube have
string-like properties; i.e.\ roughening and a L\"uscher term?
Since $\s_{coul}$ is several times greater than $\s$, the Coulomb
flux tube must be an excited state.  By what mechanism is the
string tension $\s$ of the minimal energy flux tube reduced below the value
$\s_{coul}$ of the Coulomb flux tube?   Does this mechanism
involve production of constituent gluons, along the lines of the
gluon-chain model \cite{gchain}, or is some other process at work?
We hope to address some of these issues in a future investigation.

%
%
\acknowledgments{%
J.G.\ is pleased to acknowledge the hospitality of the Niels Bohr
Institute during the course of this investigation.  Our research
is supported in part by the U.S. Department of Energy under Grant
No.\ DE-FG03-92ER40711 (J.G.), the Slovak Grant Agency for
Science, Grant No. 2/3106/2003 (\v{S}.O.), and the National Science
Foundation, Grant No. PHY-0099393 (D.Z.). }

%
%

\end{document}